\documentclass[
    aps,      
    prx,      
    reprint,  
    noeprint, 
    nofootinbib, 
    superscriptaddress, 
]{revtex4-2}
\usepackage{physics, siunitx, mathtools, amssymb, bm}
\usepackage[version=4]{mhchem}
\usepackage{graphicx}
\usepackage{hyperref}
\usepackage[capitalize]{cleveref}

\DeclareMathOperator*{\argmax}{arg\,max}
\DeclareMathOperator*{\argmin}{arg\,min}

\newcommand{\Wsc}{W_\text{sc}}
\newcommand{\Vg}{V_\text{g}}
\newcommand{\Ic}{I_\text{c}}
\newcommand{\Icp}{I_{\text{c}+}}
\newcommand{\Icm}{I_{\text{c}-}}
\newcommand{\Icpm}{I_{\text{c}\pm}}

\begin{document}
\title{Geometric dependence of critical current magnitude and nonreciprocity in planar Josephson junctions}

\author{William F. Schiela}
\email{william.schiela@nyu.edu}
\author{Melissa Mikalsen}
\author{William M. Strickland}
\author{Javad Shabani}
\email{jshabani@nyu.edu}
\affiliation{Center for Quantum Information Physics, Department of Physics, New York University, New York, NY 10003, USA}

\date{\today}
\begin{abstract}
Planar Josephson junctions in a magnetic field exhibit the superconducting diode effect, by which the critical current magnitude
depends on the polarity of the transport current.
A number of different mechanisms for the effect have been proposed.
Here, we study symmetric, T-shaped planar Josephson junctions with semiconducting weak links in an in-plane magnetic field perpendicular to an applied current bias.
In particular, we vary the longitudinal width (i.e.\ parallel to the current) of the superconducting contacts and the voltage of an electrostatic gate.
We observe an increase in both critical current and diode efficiency with increasing contact width and relate the critical current behavior to the induced coherence length of the Andreev bound states that mediate the supercurrent flow through the junction.
We further observe a linear trend, with respect to inverse contact width, of the field at which the diode efficiency is maximized, which saturates as the contact width becomes large compared to the coherence length.
The smaller field at which the critical current is maximized additionally exhibits a strong gate dependence.
We interpret these observations in the context of multiple underlying mechanisms, including spin--orbit coupling and orbital effects.
\end{abstract}
\maketitle

\section{%
    \label{sec:introduction}%
    Introduction%
}

The superconducting diode effect---an asymmetry of certain superconducting properties with respect to supercurrent polarity---is observed ubiquitously in a variety of superconducting systems due to simultaneously broken time-reversal and space-inversion symmetries.
Several different symmetry-breaking mechanisms have been proposed which may be broadly classified as either intrinsic or extrinsic.
Interest in the effect has resurged in recent years due in part to its potential to elucidate the intrinsic or microscopic properties of a material system; however, such work has been complicated by the coexistence of various extrinsic or macroscopic symmetry-breaking mechanisms.
The SDE has recently been demonstrated in
    ordinary elemental superconductors \citep{satchell2023,hou2023};
    superconductor \citep{ando2020} and superconductor--semiconductor heterostructures and nanowires \citep{sundaresh2023};
    graphene \citep{lin2022};
    and Josephson junctions \citep{margineda2023,chen2024,lotfizadeh2024,mazur2022_diode}
    and circuits thereof such as series arrays \citep{baumgartner2021_diode,baumgartner2022_diodeDresselhaus,costa2023_diodeSignReversal},
    multi-terminal junctions and superconducting quantum interference devices \citep{coraiola2023_diode,gupta2023_diode,banerjee2023_diode,reinhardt2024}.
While broken time-reversal symmetry is most commonly due to magnetism, broken space-inversion symmetry has been variously attributed to imperfections in nanofabrication or material inhomogeneities \citep{satchell2023,hou2023,chen2024}, structural inversion asymmetry of heterostructures \citep{ando2020,sundaresh2023,lotfizadeh2024,baumgartner2021_diode,baumgartner2022_diodeDresselhaus,costa2023_diodeSignReversal,reinhardt2024}, and asymmetry in the device configuration \citep{gupta2023_diode,coraiola2023_diode}.
As a result, proposed mechanisms of the SDE typically vary by system and include
    vortex effects \citep{satchell2023,hou2023,margineda2023,chen2024,sundaresh2023},
    spin--orbit coupling \citep{ando2020,lotfizadeh2024,baumgartner2021_diode,baumgartner2022_diodeDresselhaus,costa2023_diodeSignReversal,reinhardt2024,mazur2022_diode,margineda2023,yuan&fu2022,ilic&bergeret2022,costa2023_diodeMicroscopic},
    Meissner or diamagnetic currents \citep{hou2023,davydova2022,sundaresh2023},
    orbital effects \citep{banerjee2023_diode,nakamura2024},
    finite momentum Cooper pairing \citep{yuan&fu2022,mazur2022_diode,banerjee2023_diode,davydova2022,lotfizadeh2024,pal2022_diode},
    higher harmonic content in the current--phase relation \citep{baumgartner2021_diode,baumgartner2022_diodeDresselhaus,costa2023_diodeSignReversal,reinhardt2024,coraiola2023_diode,gupta2023_diode,souto2022},
    or combinations thereof.

Planar Josephson junctions with semiconducting weak links are promising platforms for scalable spintronics \citep{zutic2004review,fabian2007review} and topological superconductivity \citep{schiela2024perspective,flensberg2021review,lutchyn2018review}.
The combination of s-wave superconductivity, spin--orbit coupling, and Zeeman interaction can lead to both topological superconductivity \citep{hell2017_planarJJ,pientka2017} and the superconducting diode effect \citep{yuan&fu2022,ilic&bergeret2022}.
The longitudinal width of the superconducting contacts are an important device parameter determining the relative flux- and gate-tunability of the system topology and impacting the phenomenology of the predicted topological signatures as well as the strength of trivial orbital effects \citep{setiawan2019_narrowing,haxell2023_orbital,pekerten2024_beyondTopoJJStandardModel}.
Most experiments thus far have typically probed one of three regimes, using junctions with narrow \citep{fornieri2019}, intermediate \citep{dartiailh2021_piJump}, or wide \citep{ren2019} superconducting contacts with respect to the induced superconducting coherence length.

Here we report measurements of the superconducting diode effect in planar Josephson junctions co-fabricated on a superconductor--semiconductor heterostructure.
We study the critical current magnitude and nonreciprocity as functions of the longitudinal superconducting contact width and gate voltage in the presence of an in-plane magnetic field applied perpendicular to the current bias.
We connect the scaling of the critical current magnitude with contact width to the induced coherence length of the Andreev bound states of the junction.
In addition, the magnetic fields at which the critical current magnitude and nonreciprocity are maximized exhibit distinct behaviors with contact width and gate voltage.
We discuss possible interpretations of the data which invoke finite momentum Cooper pairing resulting from spin--orbit coupling, Meissner currents, or orbital effects.

\section{%
    \label{sec:devices-and-measurement}%
    Symmetric planar Josephson junctions%
}

\begin{figure}
    \centering
    \includegraphics[width=\linewidth]{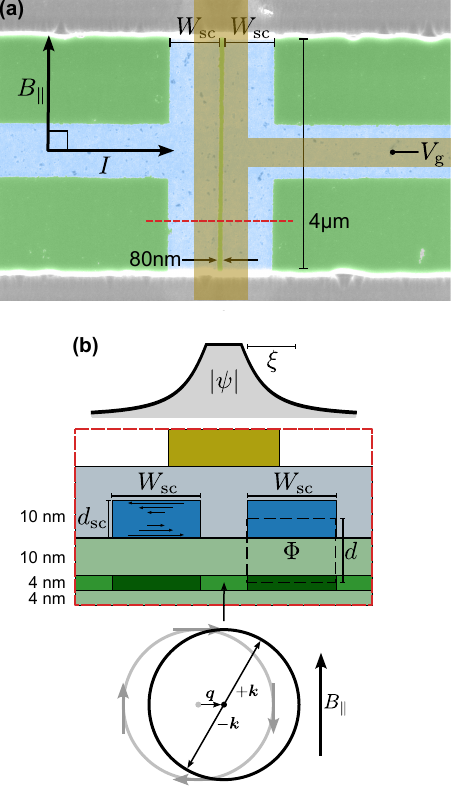}
    \caption{%
        \textbf{Symmetric T-shaped planar Josephson junctions and diode effect mechanisms.}
        \textbf{(a)} Scanning electron micrograph of a representative planar Josephson junction co-fabricated on the same chip as the devices presented here.  The superconductor (blue) and semiconductor (green) are falsely colored.  The gate (yellow) is shown schematically.  All devices are \SI{4}{\micro{m}} wide with \SI{80}{nm} separating the superconducting contacts.  Each device has a different superconducting contact width $W_\text{sc}$.  A current bias $I$ and gate voltage $\Vg$ are applied.  The in-plane magnetic field $B_\parallel$ is perpendicular to the current.  The dashed red line indicates the cross section shown schematically in (b).
        \textbf{(b)} Schematic of the device cross section indicated by the dashed red line in (a).  The magnitude of the Andreev bound state wave function $\abs{\psi}$ decays with characteristic length $\xi$ into the superconductor.  A magnetic flux $\Phi$ threads the area $W_\text{sc}d$ between the superconducting contacts (blue) and the proximitized 2DEG (dark green).  A Meissner current flows in the superconducting contacts of thickness $d_\text{sc}$.  In the 2DEG, the Rashba spin--orbit split Fermi surface (grey) is Zeeman-shifted by the in-plane magnetic field, yielding Cooper pairs of electrons with wavevectors $\bf{q}\pm\bf{k}$ (black).  (The inner Fermi surface is not shown.)
    }
    \label{fig:device}
\end{figure}

We study symmetric planar Josephson junctions with various superconducting contact widths $W_\text{sc}$, fabricated in a superconductor--semiconductor heterostructure, in the presence of an in-plane magnetic field $B_\parallel$ applied perpendicular to the current $I$, as shown in \cref{fig:device}(a).
All devices presented here were co-fabricated on a single chip within a \SI{1.6}{mm}$\times$\SI{1.6}{mm} area.

The heterostructure comprises a near-surface (In,Ga)As/InAs/(In,Ga)As (\SI{4}{nm}/\SI{4}{nm}/\SI{10}{nm}) quantum well capped by a layer of Al grown in-situ by molecular beam epitaxy, as shown in \cref{fig:device}(b).
An Al thickness of \SI{10}{nm} was inferred from the critical in-plane field of \SI{2.7}{T} \citep{meservey&tedrow1971} (see
Section S.I).
All of the junctions have a width (transverse to the current) of \SI{4}{\micro m} defined by a deep wet etch through the III--V semiconducting layers.
The \SI{80}{nm} contact separation and variable contact width $\Wsc$ were defined by selectively wet etching the Al and confirmed by scanning electron micrographs of nominally identical proxy devices co-fabricated on the same chip (see
Section S.II).
Subsequently, a \SI{60}{nm} AlOx dielectric was grown at \SI{150}{\celsius} by atomic layer deposition, followed by the deposition of \SI{10}{nm}/\SI{90}{nm} of Ti/Au gate electrodes by electron beam evaporation.

To eliminate as much as possible any nonreciprocal effects associated with an out-of-plane magnetic field component, we measure Fraunhofer interference in a narrow out-of-plane field range about the extrema of the central lobe.  The positive (negative) critical current is then obtained as the maximum (minimum) of the positive (negative) branch of the Fraunhofer pattern at each in-plane field and gate voltage setting; see
Section S.III
for details.

\section{%
    \label{sec:abs}%
    Andreev bound state coherence length%
}

\begin{figure}[tbp]
    \centering
    \includegraphics[width=\linewidth]{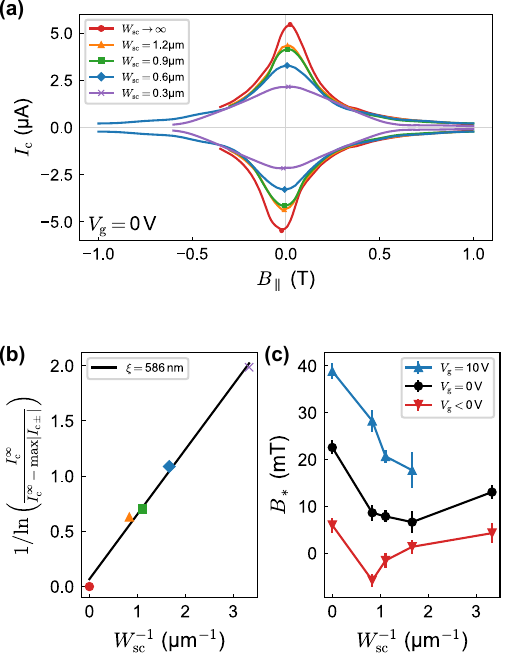}
    \caption[Critical current and coherence length]{%
        \textbf{Critical current and coherence length.}
        \textbf{(a)} Critical current $I_{\text{c}\pm}$ versus in-plane magnetic field $B_\parallel$ perpendicular to the current, for five devices with different contact widths $W_\text{sc}$, at gate voltage $V_\text{g}=0$ (see
        Section S.IV
        for similar field traces at positive and negative gate voltages).  Markers identify the extrema plotted in (b,c).
        \textbf{(b)} Extrema of $I_{\text{c}\pm}$ versus inverse superconducting contact width $W_\text{sc}^{-1}$, fit to \cref{eq:ic-vs-wsc}.  The markers correspond to the legend in (a).
        \textbf{(c)} In-plane magnetic field $B_*=\argmax_{B_\parallel}I_{\text{c}+}=-\argmin_{B_\parallel}I_{\text{c}-}$ at which $I_{\text{c}\pm}$ are extremized.  Black dots correspond to the $\Vg=0$ extrema marked in (a); blue (red) upward-pointing (downward-pointing) triangles correspond to the extrema at positive (negative) gate voltage; see
        Section S.IV.
    }%
    \label{fig:critical-current}
\end{figure}

The supercurrent through the junction is carried by Andreev bound states that are localized within the weak link and decay exponentially into the superconductor over a characteristic length $\xi$, the coherence length of the proximity-induced superconductivity, as shown in \cref{fig:device}(b) \citep{nazarov,prada2020review}.
The Andreev bound state coherence length \citep{prada2020review} $\xi_\text{ABS}=\xi/\sqrt{\tau}\abs{\sin(\varphi/2)}$ approaches $\xi$ at the critical current in the high-transparency limit $\tau\to 1$; therefore, we do not distinguish between $\xi$ and $\xi_\text{ABS}$ in the following.
If $\Wsc \lesssim \xi$, a Cooper pair may be reflected at the rear interface of the superconducting contacts, reducing the supercurrent carried by the Andreev bound state.
The probability of reflection is proportional to the amplitude $\psi\propto\exp(-\Wsc/\xi)$ of the wave function attained at that interface,
such that the critical current of the junction is suppressed by finite $\Wsc$ as
\begin{equation}
    \Ic(\Wsc) = \Ic^\infty \left(1 - e^{-W_\text{sc}/\xi}\right)
    \label{eq:ic-vs-wsc}
\end{equation}
where $\Ic^\infty = \lim_{\Wsc\to\infty} \Ic(\Wsc)$.

In \cref{fig:critical-current} we show the dependence of the critical current on superconducting contact width $\Wsc$.  \Cref{fig:critical-current}(a) shows the positive and negative critical currents $\Icpm$ as a function of in-plane magnetic field $B_\parallel$ perpendicular to the current for five devices with different $W_\text{sc}$.  The data are antisymmetric with respect to in-plane field, $\Icp(B_\parallel) = -\Icm(-B_\parallel)$, and a clear trend of increasing critical current with increasing $W_\text{sc}$ is apparent.  In \cref{fig:critical-current}(b) we plot a fit of the extrema $\max \Icp(B_\parallel) = -\min \Icm(B_\parallel) \equiv \Ic(\Wsc)$ to the model \cref{eq:ic-vs-wsc}, taking the average of the values obtained from the positive and negative branches.  The fit yields an induced coherence length $\xi=\SI{586}{nm}$.  This agrees well with the dirty limit coherence length \citep{kulik&omelyanchuk1975,mayer2019_proximitizedInAs} $\xi = \sqrt{\xi_\text{BCS}\ell}\approx\SI{570}{nm}$
where the BCS coherence length $\xi_\text{BCS} = \hbar v_\text{F} / \pi\Delta \approx \SI{2.5}{\micro m}$ with Fermi velocity $v_\text{F} = \hbar\sqrt{2\pi n}/m^* \approx \SI{3e6}{m/s}$ and induced gap $\Delta=1.764k_\text{B}T_\text{c}\approx\SI{250}{\micro eV}$.  Here we have used the density $n=\SI{6e12}{cm^{-2}}$ and mean free path $\ell=\SI{130}{nm}$ obtained from magnetotransport data, effective band mass $m^* = 0.024 m_e$ of InAs \citep{adachi_semiconductors}, and critical temperature $T_\text{c}=\SI{1.65}{K}$ measured during multiple cooldowns; see
Section S.I.
The measured coherence length is also consistent with nonlocal spectroscopic signatures of extended Andreev bound states \citep{poschl2022_1} as well as signatures of Andreev bound state hybridization across a common electrode \citep{haxell2023_andreevMolecule} in similar Al/InAs heterostructures.

Multiple works have posited mechanisms of the superconducting diode effect based on spin--orbit coupling (see e.g. Refs.~\citenum{ando2020,baumgartner2021_diode,yuan&fu2022,ilic&bergeret2022,costa2023_diodeMicroscopic,lotfizadeh2024}).  Based on these, the nonreciprocity of the critical current should be gate-tunable, as the electric field from an applied gate voltage modulates the Rashba spin--orbit coupling strength \citep{farzaneh2024}.
Following Ref.~\citenum{lotfizadeh2024}, the field $B_*=\argmax_{B_\parallel}\Icp=-\argmin_{B_\parallel}\Icm$ at which the nonreciprocal critical current $\Icpm$ is extremized is given by
\begin{equation}
    B_* \approx (1-\tau)^{1/4} \frac{k_\text{so}}{k_\text{F}} \frac{4E_\text{T}}{g^*\mu_\text{B}}.
    \label{eq:bstar}
\end{equation}
As all devices are co-fabricated on the same chip with identical contact separations of \SI{80}{nm}, we do not expect the Thouless energy $E_\text{T}$ or effective g-factor $g^*$ to vary significantly across devices.  However, we note that \cref{eq:bstar} was derived in the limit $\Wsc\to\infty$; therefore, we do not attempt to make comparisons of $B_*$ across devices with different $\Wsc$.  For a given $\Wsc$, on the other hand, increasing gate voltage increases $B_*$ via the spin--orbit wavenumber $k_\text{so}=\alpha m^*/\hbar^2$, where $\alpha$ is the Rashba spin--orbit coupling strength and $m^*$ is the effective band mass.
A changing gate voltage may also impact $B_*$ via the transparency $\tau$.
We note that the trends of increasing $B_*$ with increasing $\alpha$ and decreasing $\tau$ predicted by \cref{eq:bstar} are consistent with the results of Ref.~\citenum{costa2023_diodeMicroscopic}.

\begin{figure}[tbp]
    \centering
    \includegraphics{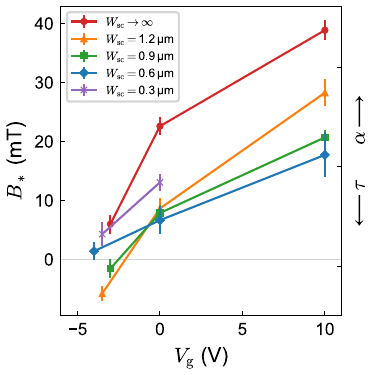}
    \caption[Gate dependence and spin--orbit coupling]{%
        \textbf{Gate dependence and spin--orbit coupling.}
        In-plane magnetic field $B_*=\argmax_{B_\parallel}\Icp=-\argmin_{B_\parallel}\Icm$ at which $I_{\text{c}\pm}$ is extremized, for five devices with different $W_\text{sc}$.  This is the same data as in \cref{fig:critical-current}(c) plotted here as a function of gate voltage $\Vg$.  The right-hand axis indicates the directions of increasing Rashba spin--orbit coupling strength $\alpha$ and junction transparency $\tau$ according to \cref{eq:bstar}.
    }%
    \label{fig:gate}
\end{figure}

In \cref{fig:critical-current}(c) we show the $\Wsc$-dependence of the extremal field $B_*$, extracted from low-field fits to Eq.~(1) of Ref.~\citenum{lotfizadeh2024}, at different gate voltages.  A nontrivial dependence on both $\Wsc$ and gate voltage is observed.  \Cref{fig:gate} shows the same data as a function of gate voltage.  All devices exhibit similar behavior in which $B_*$ increases with increasing gate voltage.  This behavior is consistent with both increasing spin--orbit coupling strength and decreasing transparency as the gate voltage increases, according to \cref{eq:bstar}.  The Rashba spin--orbit coupling strength increases with increasing gate voltage due to the increased structural inversion asymmetry of the quantum well \citep{farzaneh2024}.



\section{%
    \label{sec:diode}%
    Superconducting diode effect%
}

Planar Josephson junctions in an applied in-plane magnetic field exhibit critical current nonreciprocity, a facet of the superconducting diode effect, due to a number of mechanisms including, as shown in \cref{fig:device}(b),
Meissner screening currents \citep{davydova2022}, orbital effects \citep{banerjee2023_diode}, and the interplay of spin--orbit coupling and Zeeman interaction \citep{baumgartner2021_diode,yuan&fu2022,ilic&bergeret2022,costa2023_diodeMicroscopic,lotfizadeh2024}, each resulting in a superconducting condensate with finite momentum.
The nonreciprocity appears with the application of an in-plane magnetic field $B_\parallel$ perpendicular to the current, which is associated with a magnetic flux $\Phi$ threading the area $\Wsc d$ between the superconducting contacts and the proximitized two dimensional electron gas,
\begin{equation}
    B_\parallel \Wsc d = \Phi,
    \label{eq:flux}
\end{equation}
where $d$ is their effective separation, as shown in \cref{fig:device}(b).
We quantify the SDE using the diode efficiency
\begin{equation}
    \eta = \frac{
        \Icp - \abs{\Icm}
    }{
        \Icp + \abs{\Icm}
    }
    \label{eq:diode-efficiency}
\end{equation}
defined such that $-1 < \eta < 1$ with the sign indicating the diode polarity.

\begin{figure}[tbp]
    \centering
    \includegraphics[width=\linewidth]{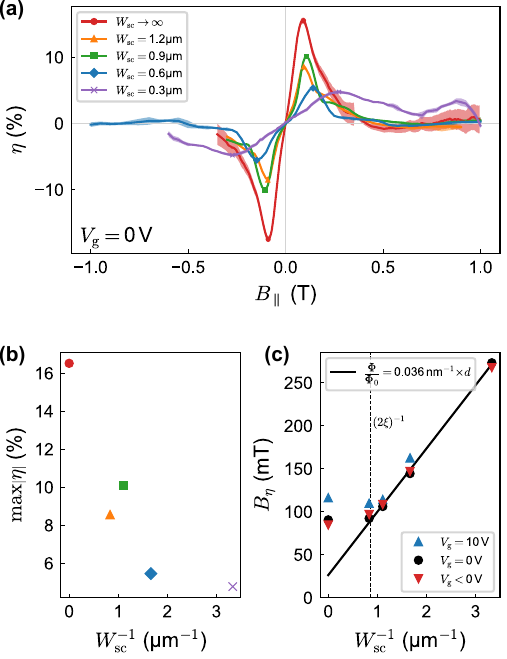}
    \caption[Diode efficiency and orbital effect]{%
        \textbf{Diode efficiency and orbital effect.}
        \textbf{(a)} Diode efficiency $\eta$, as defined in \cref{eq:diode-efficiency}, determined from the data in \cref{fig:critical-current}(a) (see
        Section S.IV
        for similar field traces at positive and negative gate voltages).  Markers identify the extrema plotted in (b,c).
        \textbf{(b)} Extrema $\max\eta=-\min\eta$ versus inverse superconducting contact width $W_\text{sc}^{-1}$.  The markers correspond to the legend in (a).
        \textbf{(c)} In-plane magnetic field $B_\eta=\argmax_{B_\parallel}\eta=-\argmin_{B_\parallel}\eta$ at which $\eta$ is extremized.  Black dots correspond to the extrema marked in (a); blue (red) upward-pointing (downward-pointing) triangles correspond to the extrema at positive (negative) gate voltage.  The solid black line is a fit to \cref{eq:flux} of the $V_\text{g}=0$ series, excluding the point at $W_\text{sc}^{-1}=0$.
    }%
    \label{fig:diode-efficiency}
\end{figure}

\Cref{fig:diode-efficiency}(a) shows the diode efficiency obtained from the curves in \cref{fig:critical-current}(a).
All devices exhibit similar behavior: $\eta(B_\parallel)$ is antisymmetric with respect to field, $\eta(B_\parallel)=-\eta(-B_\parallel)$, as expected from the antisymmetry of $\Icpm(B_\parallel)$.
Likewise for all devices $\eta(B_\parallel)$ attains extrema at finite field, and all but $\Wsc=\SI{0.3}{\micro m}$ exhibit a weak sign change in the range \SIrange{400}{600}{mT}.
These features---antisymmetry, extrema, and finite-field sign changes---are qualitatively consistent with multiple SDE mechanisms including orbital effects \citep{banerjee2023_diode,nakamura2024}, diamagnetic currents \citep{sundaresh2023}, and simultaneous spin--orbit and Zeeman interactions \citep{baumgartner2021_diode,yuan&fu2022,costa2023_diodeMicroscopic,costa2023_diodeSignReversal,lotfizadeh2024}.

We now consider the dependence of the diode efficiency on $\Wsc$.
\Cref{fig:diode-efficiency}(b) shows a general trend of increasing peak efficiency $\max(\eta)=-\min(\eta)$ with increasing $\Wsc$, although the behavior does not follow a clear exponential trend like the extrema of $\Ic$ in \cref{fig:critical-current}(b).
\Cref{fig:diode-efficiency}(c) shows the in-plane field $B_\eta = \argmax_{B_\parallel}\eta = -\argmin_{B_\parallel}\eta$ at which the diode efficiency is extremized at different gate voltages.  A strong dependence on $\Wsc$ is observed, as compared to the much weaker dependence on gate voltage.
A fit to \cref{eq:flux} yields $\Phi/\Phi_0\approx\numrange{0.36}{0.61}$ for $d=\SIrange{10}{17}{nm}$, corresponding to an effective Al--2DEG separation that either excludes or includes half of the Al and InAs quantum well thicknesses (see \cref{fig:device}(b)).
The deviation from this linear trend for $\Wsc\to\infty$ may be due to length-limiting effects such as the formation of Josephson vortices between the Al and InAs layers \citep{sundaresh2023}.
We note that $B_\eta$ saturates for $\Wsc\gtrsim 2\xi\approx\SI{1.15}{\micro m}$.
These observations are consistent with an orbital mechanism by which the extremum of $\eta$ should occur before the first sign change at $\Phi=\Phi_0$; we do not, however, observe these sign changes at $\Phi=\Phi_0$, nor do we observe the expected $\Phi_0$-periodic oscillation of $\eta$ at higher fields \citep{banerjee2023_diode}.
Instead, the sign changes are observed at a significantly larger flux $\Phi > \Phi_0$, above which any oscillations are damped out, likely due to the collapse of the induced gap at large magnetic fields by the Zeeman energy.
The relatively weak gate dependence of $B_\eta$ shown in \cref{fig:diode-efficiency}(c) is also consistent with a spin--orbit coupling based mechanism, where a weak dependence of $B_\eta$ on spin--orbit coupling strength is in fact expected \citep{costa2023_diodeMicroscopic}.
Furthermore, while most theory is done in the limit $W_\text{sc}\to\infty$ in which the only relevant device length scale is the distance $W_\text{n}$ between the superconducting contacts, one can expect some $\Wsc$ dependence for $\Wsc \lesssim \xi$ for any mechanism yielding finite Cooper pair momentum $2q$ due to the resonance condition at the system boundaries, $2q(2\Wsc + W_\text{n}) = (2n+1)\pi/2$ for integer $n$, which identifies the zeroes of the diode efficiency \citep{hart2017,lotfizadeh2024}.

Generically, a single-mode Josephson junction comprising two superconductors with finite Cooper pair momentum $2q$ along the transport direction is expected to exhibit a maximum critical current nonreciprocity at the value $q_0$ satisfying \citep{davydova2022,scharf2024}\footnotemark{}
\footnotetext{We note that this expression differs from that of \citet{davydova2022} but follows from the maximization of Eq.~13 of Ref.~\citenum{davydova2022} with respect to $q$ and agrees with the results of \citet{scharf2024}.}
\begin{equation}
    \hbar q_0 v_\text{F} =
    \frac
    {\Delta}
    {\sqrt{1+\left(\frac{\pi}{4}\right)^2}} \approx 0.8\Delta,
    \label{eq:argmax-expression}
\end{equation}
independent of the microscopic mechanism responsible for $q$.
By the Meissner screening mechanism \citep{davydova2022}, a finite $q_\text{M}=(e/\hbar)B_\parallel\lambda_\text{L}$ in the proximitized 2DEG is inherited from the Meissner screening current induced at the lower surface of the parent superconductor by the magnetic field (\cref{fig:device}(b)).  For a superconducting film of thickness $d_\text{sc}$ thinner than the London penetration depth $\lambda_\text{L}$, we take $\lambda_\text{L}\to d_\text{sc}/2$ to estimate $q_\text{M}/B_\parallel\approx\SI{7.6e6}{m^{-1}T^{-1}}$ for $d_\text{sc}=\SI{10}{nm}$, with an expected maximum according to \cref{eq:argmax-expression} at $B_\parallel^\text{max}\approx\SI{13}{mT}$ using the same parameters as in \cref{sec:abs}.  Similarly, the orbital effect \citep{banerjee2023_diode} of the magnetic field on electrons tunneling between the parent superconductor and the 2DEG induces in the 2DEG a finite $q_\text{orb}=(e/\hbar)B_\parallel d$, from which we estimate $q_\text{orb}/B_\parallel\approx\SI{2.6e7}{m^{-1}T^{-1}}$ and $B_\parallel^\text{max}\approx\SI{4}{mT}$.  Finally, in the limit of strong spin--orbit coupling ($\alpha k_\text{F} \gg E_\text{Z}$), the Zeeman interaction shifts the center of the (outer) Rashba spin--orbit split Fermi surface to finite $q_\text{RZ} = E_\text{Z}/\hbar v_\text{F}$ \citep{hart2017,setiawan2019_narrowing,lotfizadeh2024} where $E_\text{Z}=g^*\mu_\text{B}B_\parallel/2$ is the Zeeman energy, implying $q_\text{RZ}/B_\parallel\approx\SI{1.5e5}{m^{-1}T^{-1}}$ and $B_\parallel^\text{max}\approx\SI{690}{mT}$ for an effective $\abs{g^*}=10$.
We note that $q_\text{orb} > q_\text{M} > q_\text{RZ}$ for a given $B_\parallel$, suggesting orbital effects are strong in our system; however, none of the above estimates of the extremal field $B_\parallel^\text{max}$ are a particularly good match to the experimental values observed in \cref{fig:diode-efficiency}(c).
We note that the estimates obtained above are linear approximations that neglect the dimension $\Wsc$ as well as the effect of kinetic inductance which is expected to be non-negligible in superconducting thin films.
Furthermore, the single-mode model leading to \cref{eq:argmax-expression} neglects the many transverse modes present in wide planar junctions.
\citet{costa2023_diodeMicroscopic} have shown that in the presence of both Rashba spin--orbit and magnetic exchange interactions, modes with larger transverse wavenumbers exhibit larger anomalous phase shifts and experience a stronger effective magnetic exchange interaction than in the one-dimensional case.
From the perspective of finite Cooper pair momentum, modes with larger transverse wavenumbers have smaller longitudinal Fermi velocities and therefore larger $q_\text{RZ} \propto v_\text{F}^{-1}$, which would increase the single-mode $q_\text{RZ}$ approximation (and decrease the corresponding $B_\parallel^\text{max}$) given above.

\section{%
    \label{sec:conclusion}%
    Conclusion%
}

We have studied the geometric effect of the superconducting contact width of symmetric planar Josephson junctions on the magnitude and nonreciprocity of critical current.
The critical current magnitude exponentially approaches a maximum value as the contact width increases, reflecting the induced superconducting coherence length of the Andreev bound states in the junction.
The maximum nonreciprocity also tends to increase for larger contact widths.
We have interpreted the data with respect to different possible mechanisms of the superconducting diode effect, primarily orbital effects and concurrent spin--orbit and Zeeman interactions, which both induce finite momentum Cooper pairs.
The contact width dependence of the field of maximum nonreciprocity invites an orbital interpretation, while the gate tunability of the smaller field of maximum critical current lends credence to a spin--orbit coupling based interpretation.
We expect both orbital and spin--orbit coupling effects coexist in our system.
Our work displays the importance of geometric effects on the phenomenology of planar Josephson junctions, as well as the utility of simple geometric modifications in device optimization and hypothesis testing.
Future experiments could alter geometric aspects of the heterostructure to further study the various potential contributions to the superconducting diode effect, for example by varying the superconductor thickness around the penetration depth, or by moving the 2DEG closer to or further away from the surface.

\section{Acknowledgments}
We would like to acknowledge Daniel Crawford, Stefan Ili\'c, F.~Sebastian Bergeret, and Alex Matos-Abiague for fruitful discussions and feedback on this manuscript.  This work was supported by ONR N00014-22-1-2764 and ONR N00014-21-1-2450.  W.~F.~S. acknowledges support from the NDSEG Fellowship.

\bibliographystyle{apsrev4-2}
\bibliography{papers_clean,textbooks_clean}

\begin{thebibliography}{47}%
\makeatletter
\providecommand \@ifxundefined [1]{%
 \@ifx{#1\undefined}
}%
\providecommand \@ifnum [1]{%
 \ifnum #1\expandafter \@firstoftwo
 \else \expandafter \@secondoftwo
 \fi
}%
\providecommand \@ifx [1]{%
 \ifx #1\expandafter \@firstoftwo
 \else \expandafter \@secondoftwo
 \fi
}%
\providecommand \natexlab [1]{#1}%
\providecommand \enquote  [1]{``#1''}%
\providecommand \bibnamefont  [1]{#1}%
\providecommand \bibfnamefont [1]{#1}%
\providecommand \citenamefont [1]{#1}%
\providecommand \href@noop [0]{\@secondoftwo}%
\providecommand \href [0]{\begingroup \@sanitize@url \@href}%
\providecommand \@href[1]{\@@startlink{#1}\@@href}%
\providecommand \@@href[1]{\endgroup#1\@@endlink}%
\providecommand \@sanitize@url [0]{\catcode `\\12\catcode `\$12\catcode
  `\&12\catcode `\#12\catcode `\^12\catcode `\_12\catcode `\%12\relax}%
\providecommand \@@startlink[1]{}%
\providecommand \@@endlink[0]{}%
\providecommand \url  [0]{\begingroup\@sanitize@url \@url }%
\providecommand \@url [1]{\endgroup\@href {#1}{\urlprefix }}%
\providecommand \urlprefix  [0]{URL }%
\providecommand \Eprint [0]{\href }%
\providecommand \doibase [0]{https://doi.org/}%
\providecommand \selectlanguage [0]{\@gobble}%
\providecommand \bibinfo  [0]{\@secondoftwo}%
\providecommand \bibfield  [0]{\@secondoftwo}%
\providecommand \translation [1]{[#1]}%
\providecommand \BibitemOpen [0]{}%
\providecommand \bibitemStop [0]{}%
\providecommand \bibitemNoStop [0]{.\EOS\space}%
\providecommand \EOS [0]{\spacefactor3000\relax}%
\providecommand \BibitemShut  [1]{\csname bibitem#1\endcsname}%
\let\auto@bib@innerbib\@empty
\bibitem [{\citenamefont {Satchell}\ \emph {et~al.}(2023)\citenamefont
  {Satchell}, \citenamefont {Shepley}, \citenamefont {Rosamond},\ and\
  \citenamefont {Burnell}}]{satchell2023}%
  \BibitemOpen
  \bibfield  {author} {\bibinfo {author} {\bibfnamefont {N.}~\bibnamefont
  {Satchell}}, \bibinfo {author} {\bibfnamefont {P.~M.}\ \bibnamefont
  {Shepley}}, \bibinfo {author} {\bibfnamefont {M.~C.}\ \bibnamefont
  {Rosamond}},\ and\ \bibinfo {author} {\bibfnamefont {G.}~\bibnamefont
  {Burnell}},\ }\href {https://doi.org/10.1063/5.0141576} {\bibfield  {journal}
  {\bibinfo  {journal} {Journal of Applied Physics}\ }\textbf {\bibinfo
  {volume} {133}},\ \bibinfo {pages} {203901} (\bibinfo {year} {2023})},\
  \bibinfo {note} {arXiv:2301.02706 [cond-mat] type: article}\BibitemShut
  {NoStop}%
\bibitem [{\citenamefont {Hou}\ \emph {et~al.}(2023)\citenamefont {Hou},
  \citenamefont {Nichele}, \citenamefont {Chi}, \citenamefont {Lodesani},
  \citenamefont {Wu}, \citenamefont {Ritter}, \citenamefont {Haxell},
  \citenamefont {Davydova}, \citenamefont {Ilić}, \citenamefont
  {Glezakou-Elbert}, \citenamefont {Varambally}, \citenamefont {Bergeret},
  \citenamefont {Kamra}, \citenamefont {Fu}, \citenamefont {Lee},\ and\
  \citenamefont {Moodera}}]{hou2023}%
  \BibitemOpen
  \bibfield  {author} {\bibinfo {author} {\bibfnamefont {Y.}~\bibnamefont
  {Hou}}, \bibinfo {author} {\bibfnamefont {F.}~\bibnamefont {Nichele}},
  \bibinfo {author} {\bibfnamefont {H.}~\bibnamefont {Chi}}, \bibinfo {author}
  {\bibfnamefont {A.}~\bibnamefont {Lodesani}}, \bibinfo {author}
  {\bibfnamefont {Y.}~\bibnamefont {Wu}}, \bibinfo {author} {\bibfnamefont
  {M.~F.}\ \bibnamefont {Ritter}}, \bibinfo {author} {\bibfnamefont {D.~Z.}\
  \bibnamefont {Haxell}}, \bibinfo {author} {\bibfnamefont {M.}~\bibnamefont
  {Davydova}}, \bibinfo {author} {\bibfnamefont {S.}~\bibnamefont {Ilić}},
  \bibinfo {author} {\bibfnamefont {O.}~\bibnamefont {Glezakou-Elbert}},
  \bibinfo {author} {\bibfnamefont {A.}~\bibnamefont {Varambally}}, \bibinfo
  {author} {\bibfnamefont {F.~S.}\ \bibnamefont {Bergeret}}, \bibinfo {author}
  {\bibfnamefont {A.}~\bibnamefont {Kamra}}, \bibinfo {author} {\bibfnamefont
  {L.}~\bibnamefont {Fu}}, \bibinfo {author} {\bibfnamefont {P.~A.}\
  \bibnamefont {Lee}},\ and\ \bibinfo {author} {\bibfnamefont {J.~S.}\
  \bibnamefont {Moodera}},\ }\href
  {https://doi.org/10.1103/PhysRevLett.131.027001} {\bibfield  {journal}
  {\bibinfo  {journal} {Physical Review Letters}\ }\textbf {\bibinfo {volume}
  {131}},\ \bibinfo {pages} {027001} (\bibinfo {year} {2023})}\BibitemShut
  {NoStop}%
\bibitem [{\citenamefont {Ando}\ \emph {et~al.}(2020)\citenamefont {Ando},
  \citenamefont {Miyasaka}, \citenamefont {Li}, \citenamefont {Ishizuka},
  \citenamefont {Arakawa}, \citenamefont {Shiota}, \citenamefont {Moriyama},
  \citenamefont {Yanase},\ and\ \citenamefont {Ono}}]{ando2020}%
  \BibitemOpen
  \bibfield  {author} {\bibinfo {author} {\bibfnamefont {F.}~\bibnamefont
  {Ando}}, \bibinfo {author} {\bibfnamefont {Y.}~\bibnamefont {Miyasaka}},
  \bibinfo {author} {\bibfnamefont {T.}~\bibnamefont {Li}}, \bibinfo {author}
  {\bibfnamefont {J.}~\bibnamefont {Ishizuka}}, \bibinfo {author}
  {\bibfnamefont {T.}~\bibnamefont {Arakawa}}, \bibinfo {author} {\bibfnamefont
  {Y.}~\bibnamefont {Shiota}}, \bibinfo {author} {\bibfnamefont
  {T.}~\bibnamefont {Moriyama}}, \bibinfo {author} {\bibfnamefont
  {Y.}~\bibnamefont {Yanase}},\ and\ \bibinfo {author} {\bibfnamefont
  {T.}~\bibnamefont {Ono}},\ }\href {https://doi.org/10.1038/s41586-020-2590-4}
  {\bibfield  {journal} {\bibinfo  {journal} {Nature}\ }\textbf {\bibinfo
  {volume} {584}},\ \bibinfo {pages} {373} (\bibinfo {year}
  {2020})}\BibitemShut {NoStop}%
\bibitem [{\citenamefont {Sundaresh}\ \emph {et~al.}(2023)\citenamefont
  {Sundaresh}, \citenamefont {Vayrynen}, \citenamefont {Lyanda-Geller},\ and\
  \citenamefont {Rokhinson}}]{sundaresh2023}%
  \BibitemOpen
  \bibfield  {author} {\bibinfo {author} {\bibfnamefont {A.}~\bibnamefont
  {Sundaresh}}, \bibinfo {author} {\bibfnamefont {J.~I.}\ \bibnamefont
  {Vayrynen}}, \bibinfo {author} {\bibfnamefont {Y.}~\bibnamefont
  {Lyanda-Geller}},\ and\ \bibinfo {author} {\bibfnamefont {L.~P.}\
  \bibnamefont {Rokhinson}},\ }\href
  {https://doi.org/10.1038/s41467-023-36786-5} {\bibfield  {journal} {\bibinfo
  {journal} {Nature Communications}\ }\textbf {\bibinfo {volume} {14}},\
  \bibinfo {pages} {1628} (\bibinfo {year} {2023})},\ \bibinfo {note}
  {arXiv:2207.03633 [cond-mat]}\BibitemShut {NoStop}%
\bibitem [{\citenamefont {Lin}\ \emph {et~al.}(2022)\citenamefont {Lin},
  \citenamefont {Siriviboon}, \citenamefont {Scammell}, \citenamefont {Liu},
  \citenamefont {Rhodes}, \citenamefont {Watanabe}, \citenamefont {Taniguchi},
  \citenamefont {Hone}, \citenamefont {Scheurer},\ and\ \citenamefont
  {Li}}]{lin2022}%
  \BibitemOpen
  \bibfield  {author} {\bibinfo {author} {\bibfnamefont {J.-X.}\ \bibnamefont
  {Lin}}, \bibinfo {author} {\bibfnamefont {P.}~\bibnamefont {Siriviboon}},
  \bibinfo {author} {\bibfnamefont {H.~D.}\ \bibnamefont {Scammell}}, \bibinfo
  {author} {\bibfnamefont {S.}~\bibnamefont {Liu}}, \bibinfo {author}
  {\bibfnamefont {D.}~\bibnamefont {Rhodes}}, \bibinfo {author} {\bibfnamefont
  {K.}~\bibnamefont {Watanabe}}, \bibinfo {author} {\bibfnamefont
  {T.}~\bibnamefont {Taniguchi}}, \bibinfo {author} {\bibfnamefont
  {J.}~\bibnamefont {Hone}}, \bibinfo {author} {\bibfnamefont {M.~S.}\
  \bibnamefont {Scheurer}},\ and\ \bibinfo {author} {\bibfnamefont {J.~I.~A.}\
  \bibnamefont {Li}},\ }\href {https://doi.org/10.1038/s41567-022-01700-1}
  {\bibfield  {journal} {\bibinfo  {journal} {Nature Physics}\ }\textbf
  {\bibinfo {volume} {18}},\ \bibinfo {pages} {1221} (\bibinfo {year}
  {2022})}\BibitemShut {NoStop}%
\bibitem [{\citenamefont {Margineda}\ \emph {et~al.}(2023)\citenamefont
  {Margineda}, \citenamefont {Crippa}, \citenamefont {Strambini}, \citenamefont
  {Fukaya}, \citenamefont {Mercaldo}, \citenamefont {Cuoco},\ and\
  \citenamefont {Giazotto}}]{margineda2023}%
  \BibitemOpen
  \bibfield  {author} {\bibinfo {author} {\bibfnamefont {D.}~\bibnamefont
  {Margineda}}, \bibinfo {author} {\bibfnamefont {A.}~\bibnamefont {Crippa}},
  \bibinfo {author} {\bibfnamefont {E.}~\bibnamefont {Strambini}}, \bibinfo
  {author} {\bibfnamefont {Y.}~\bibnamefont {Fukaya}}, \bibinfo {author}
  {\bibfnamefont {M.~T.}\ \bibnamefont {Mercaldo}}, \bibinfo {author}
  {\bibfnamefont {M.}~\bibnamefont {Cuoco}},\ and\ \bibinfo {author}
  {\bibfnamefont {F.}~\bibnamefont {Giazotto}},\ }\href
  {https://doi.org/10.1038/s42005-023-01458-9} {\bibfield  {journal} {\bibinfo
  {journal} {Communications Physics}\ }\textbf {\bibinfo {volume} {6}},\
  \bibinfo {pages} {1} (\bibinfo {year} {2023})}\BibitemShut {NoStop}%
\bibitem [{\citenamefont {Chen}\ \emph {et~al.}(2024)\citenamefont {Chen},
  \citenamefont {Park}, \citenamefont {Vool}, \citenamefont {Maksimovic},
  \citenamefont {Broadway}, \citenamefont {Flaks}, \citenamefont {Zhou},
  \citenamefont {Maletinsky}, \citenamefont {Stern},\ and\ \citenamefont
  {Yacoby}}]{chen2024}%
  \BibitemOpen
  \bibfield  {author} {\bibinfo {author} {\bibfnamefont {S.}~\bibnamefont
  {Chen}}, \bibinfo {author} {\bibfnamefont {S.}~\bibnamefont {Park}}, \bibinfo
  {author} {\bibfnamefont {U.}~\bibnamefont {Vool}}, \bibinfo {author}
  {\bibfnamefont {N.}~\bibnamefont {Maksimovic}}, \bibinfo {author}
  {\bibfnamefont {D.~A.}\ \bibnamefont {Broadway}}, \bibinfo {author}
  {\bibfnamefont {M.}~\bibnamefont {Flaks}}, \bibinfo {author} {\bibfnamefont
  {T.~X.}\ \bibnamefont {Zhou}}, \bibinfo {author} {\bibfnamefont
  {P.}~\bibnamefont {Maletinsky}}, \bibinfo {author} {\bibfnamefont
  {A.}~\bibnamefont {Stern}},\ and\ \bibinfo {author} {\bibfnamefont {B.~I.
  H.~A.}\ \bibnamefont {Yacoby}}\ }\href
  {https://doi.org/10.48550/arXiv.2402.02472} {10.48550/arXiv.2402.02472}
  (\bibinfo {year} {2024}),\ \bibinfo {note} {arXiv:2402.02472 [cond-mat] type:
  article}\BibitemShut {NoStop}%
\bibitem [{\citenamefont {Lotfizadeh}\ \emph {et~al.}(2024)\citenamefont
  {Lotfizadeh}, \citenamefont {Schiela}, \citenamefont {Pekerten},
  \citenamefont {Yu}, \citenamefont {Elfeky}, \citenamefont {Strickland},
  \citenamefont {Matos-Abiague},\ and\ \citenamefont
  {Shabani}}]{lotfizadeh2024}%
  \BibitemOpen
  \bibfield  {author} {\bibinfo {author} {\bibfnamefont {N.}~\bibnamefont
  {Lotfizadeh}}, \bibinfo {author} {\bibfnamefont {W.~F.}\ \bibnamefont
  {Schiela}}, \bibinfo {author} {\bibfnamefont {B.}~\bibnamefont {Pekerten}},
  \bibinfo {author} {\bibfnamefont {P.}~\bibnamefont {Yu}}, \bibinfo {author}
  {\bibfnamefont {B.~H.}\ \bibnamefont {Elfeky}}, \bibinfo {author}
  {\bibfnamefont {W.~M.}\ \bibnamefont {Strickland}}, \bibinfo {author}
  {\bibfnamefont {A.}~\bibnamefont {Matos-Abiague}},\ and\ \bibinfo {author}
  {\bibfnamefont {J.}~\bibnamefont {Shabani}},\ }\href
  {https://doi.org/10.1038/s42005-024-01618-5} {\bibfield  {journal} {\bibinfo
  {journal} {Communications Physics}\ }\textbf {\bibinfo {volume} {7}},\
  \bibinfo {pages} {1} (\bibinfo {year} {2024})},\ \bibinfo {note}
  {arXiv:2303.01902 [cond-mat] type: article}\BibitemShut {NoStop}%
\bibitem [{\citenamefont {Mazur}\ \emph {et~al.}(2022)\citenamefont {Mazur},
  \citenamefont {van Loo}, \citenamefont {van Driel}, \citenamefont {Wang},
  \citenamefont {Badawy}, \citenamefont {Gazibegovic}, \citenamefont
  {Bakkers},\ and\ \citenamefont {Kouwenhoven}}]{mazur2022_diode}%
  \BibitemOpen
  \bibfield  {author} {\bibinfo {author} {\bibfnamefont {G.~P.}\ \bibnamefont
  {Mazur}}, \bibinfo {author} {\bibfnamefont {N.}~\bibnamefont {van Loo}},
  \bibinfo {author} {\bibfnamefont {D.}~\bibnamefont {van Driel}}, \bibinfo
  {author} {\bibfnamefont {J.-Y.}\ \bibnamefont {Wang}}, \bibinfo {author}
  {\bibfnamefont {G.}~\bibnamefont {Badawy}}, \bibinfo {author} {\bibfnamefont
  {S.}~\bibnamefont {Gazibegovic}}, \bibinfo {author} {\bibfnamefont {E.~P.
  A.~M.}\ \bibnamefont {Bakkers}},\ and\ \bibinfo {author} {\bibfnamefont
  {L.~P.}\ \bibnamefont {Kouwenhoven}}\ }\href
  {https://doi.org/10.48550/arXiv.2211.14283} {10.48550/arXiv.2211.14283}
  (\bibinfo {year} {2022}),\ \bibinfo {note} {arXiv:2211.14283 [cond-mat] type:
  article}\BibitemShut {NoStop}%
\bibitem [{\citenamefont {Baumgartner}\ \emph {et~al.}(2021)\citenamefont
  {Baumgartner}, \citenamefont {Fuchs}, \citenamefont {Costa}, \citenamefont
  {Reinhardt}, \citenamefont {Gronin}, \citenamefont {Gardner}, \citenamefont
  {Lindemann}, \citenamefont {Manfra}, \citenamefont {Junior}, \citenamefont
  {Kochan}, \citenamefont {Fabian}, \citenamefont {Paradiso},\ and\
  \citenamefont {Strunk}}]{baumgartner2021_diode}%
  \BibitemOpen
  \bibfield  {author} {\bibinfo {author} {\bibfnamefont {C.}~\bibnamefont
  {Baumgartner}}, \bibinfo {author} {\bibfnamefont {L.}~\bibnamefont {Fuchs}},
  \bibinfo {author} {\bibfnamefont {A.}~\bibnamefont {Costa}}, \bibinfo
  {author} {\bibfnamefont {S.}~\bibnamefont {Reinhardt}}, \bibinfo {author}
  {\bibfnamefont {S.}~\bibnamefont {Gronin}}, \bibinfo {author} {\bibfnamefont
  {G.~C.}\ \bibnamefont {Gardner}}, \bibinfo {author} {\bibfnamefont
  {T.}~\bibnamefont {Lindemann}}, \bibinfo {author} {\bibfnamefont {M.~J.}\
  \bibnamefont {Manfra}}, \bibinfo {author} {\bibfnamefont {P.~E.~F.}\
  \bibnamefont {Junior}}, \bibinfo {author} {\bibfnamefont {D.}~\bibnamefont
  {Kochan}}, \bibinfo {author} {\bibfnamefont {J.}~\bibnamefont {Fabian}},
  \bibinfo {author} {\bibfnamefont {N.}~\bibnamefont {Paradiso}},\ and\
  \bibinfo {author} {\bibfnamefont {C.}~\bibnamefont {Strunk}},\ }\href
  {https://doi.org/10.1038/s41565-021-01009-9} {\bibfield  {journal} {\bibinfo
  {journal} {Nature Nanotechnology}\ }\textbf {\bibinfo {volume} {17}},\
  \bibinfo {pages} {39} (\bibinfo {year} {2021})}\BibitemShut {NoStop}%
\bibitem [{\citenamefont {Baumgartner}\ \emph {et~al.}(2022)\citenamefont
  {Baumgartner}, \citenamefont {Fuchs}, \citenamefont {Costa}, \citenamefont
  {Pic{\'{o}}-Cort{\'{e}}s}, \citenamefont {Reinhardt}, \citenamefont {Gronin},
  \citenamefont {Gardner}, \citenamefont {Lindemann}, \citenamefont {Manfra},
  \citenamefont {Junior}, \citenamefont {Kochan}, \citenamefont {Fabian},
  \citenamefont {Paradiso},\ and\ \citenamefont
  {Strunk}}]{baumgartner2022_diodeDresselhaus}%
  \BibitemOpen
  \bibfield  {author} {\bibinfo {author} {\bibfnamefont {C.}~\bibnamefont
  {Baumgartner}}, \bibinfo {author} {\bibfnamefont {L.}~\bibnamefont {Fuchs}},
  \bibinfo {author} {\bibfnamefont {A.}~\bibnamefont {Costa}}, \bibinfo
  {author} {\bibfnamefont {J.}~\bibnamefont {Pic{\'{o}}-Cort{\'{e}}s}},
  \bibinfo {author} {\bibfnamefont {S.}~\bibnamefont {Reinhardt}}, \bibinfo
  {author} {\bibfnamefont {S.}~\bibnamefont {Gronin}}, \bibinfo {author}
  {\bibfnamefont {G.~C.}\ \bibnamefont {Gardner}}, \bibinfo {author}
  {\bibfnamefont {T.}~\bibnamefont {Lindemann}}, \bibinfo {author}
  {\bibfnamefont {M.~J.}\ \bibnamefont {Manfra}}, \bibinfo {author}
  {\bibfnamefont {P.~E.~F.}\ \bibnamefont {Junior}}, \bibinfo {author}
  {\bibfnamefont {D.}~\bibnamefont {Kochan}}, \bibinfo {author} {\bibfnamefont
  {J.}~\bibnamefont {Fabian}}, \bibinfo {author} {\bibfnamefont
  {N.}~\bibnamefont {Paradiso}},\ and\ \bibinfo {author} {\bibfnamefont
  {C.}~\bibnamefont {Strunk}},\ }\href
  {https://doi.org/10.1088/1361-648x/ac4d5e} {\bibfield  {journal} {\bibinfo
  {journal} {Journal of Physics: Condensed Matter}\ }\textbf {\bibinfo {volume}
  {34}},\ \bibinfo {pages} {154005} (\bibinfo {year} {2022})}\BibitemShut
  {NoStop}%
\bibitem [{\citenamefont {Costa}\ \emph
  {et~al.}(2023{\natexlab{a}})\citenamefont {Costa}, \citenamefont
  {Baumgartner}, \citenamefont {Reinhardt}, \citenamefont {Berger},
  \citenamefont {Gronin}, \citenamefont {Gardner}, \citenamefont {Lindemann},
  \citenamefont {Manfra}, \citenamefont {Kochan}, \citenamefont {Fabian},
  \citenamefont {Paradiso},\ and\ \citenamefont
  {Strunk}}]{costa2023_diodeSignReversal}%
  \BibitemOpen
  \bibfield  {author} {\bibinfo {author} {\bibfnamefont {A.}~\bibnamefont
  {Costa}}, \bibinfo {author} {\bibfnamefont {C.}~\bibnamefont {Baumgartner}},
  \bibinfo {author} {\bibfnamefont {S.}~\bibnamefont {Reinhardt}}, \bibinfo
  {author} {\bibfnamefont {J.}~\bibnamefont {Berger}}, \bibinfo {author}
  {\bibfnamefont {S.}~\bibnamefont {Gronin}}, \bibinfo {author} {\bibfnamefont
  {G.~C.}\ \bibnamefont {Gardner}}, \bibinfo {author} {\bibfnamefont
  {T.}~\bibnamefont {Lindemann}}, \bibinfo {author} {\bibfnamefont {M.~J.}\
  \bibnamefont {Manfra}}, \bibinfo {author} {\bibfnamefont {D.}~\bibnamefont
  {Kochan}}, \bibinfo {author} {\bibfnamefont {J.}~\bibnamefont {Fabian}},
  \bibinfo {author} {\bibfnamefont {N.}~\bibnamefont {Paradiso}},\ and\
  \bibinfo {author} {\bibfnamefont {C.}~\bibnamefont {Strunk}},\ }\bibfield
  {journal} {\bibinfo  {journal} {Nature Nanotechnology}\ }\href
  {https://doi.org/10.1038/s41565-023-01451-x} {10.1038/s41565-023-01451-x}
  (\bibinfo {year} {2023}{\natexlab{a}}),\ \bibinfo {note} {arXiv:2212.13460
  [cond-mat]}\BibitemShut {NoStop}%
\bibitem [{\citenamefont {Coraiola}\ \emph {et~al.}(2023)\citenamefont
  {Coraiola}, \citenamefont {Svetogorov}, \citenamefont {Haxell}, \citenamefont
  {Sabonis}, \citenamefont {Hinderling}, \citenamefont {Kate}, \citenamefont
  {Cheah}, \citenamefont {Krizek}, \citenamefont {Schott}, \citenamefont
  {Wegscheider}, \citenamefont {Cuevas}, \citenamefont {Belzig},\ and\
  \citenamefont {Nichele}}]{coraiola2023_diode}%
  \BibitemOpen
  \bibfield  {author} {\bibinfo {author} {\bibfnamefont {M.}~\bibnamefont
  {Coraiola}}, \bibinfo {author} {\bibfnamefont {A.~E.}\ \bibnamefont
  {Svetogorov}}, \bibinfo {author} {\bibfnamefont {D.~Z.}\ \bibnamefont
  {Haxell}}, \bibinfo {author} {\bibfnamefont {D.}~\bibnamefont {Sabonis}},
  \bibinfo {author} {\bibfnamefont {M.}~\bibnamefont {Hinderling}}, \bibinfo
  {author} {\bibfnamefont {S.~C.~t.}\ \bibnamefont {Kate}}, \bibinfo {author}
  {\bibfnamefont {E.}~\bibnamefont {Cheah}}, \bibinfo {author} {\bibfnamefont
  {F.}~\bibnamefont {Krizek}}, \bibinfo {author} {\bibfnamefont
  {R.}~\bibnamefont {Schott}}, \bibinfo {author} {\bibfnamefont
  {W.}~\bibnamefont {Wegscheider}}, \bibinfo {author} {\bibfnamefont {J.~C.}\
  \bibnamefont {Cuevas}}, \bibinfo {author} {\bibfnamefont {W.}~\bibnamefont
  {Belzig}},\ and\ \bibinfo {author} {\bibfnamefont {F.}~\bibnamefont
  {Nichele}}\ }\href {https://doi.org/10.48550/arXiv.2312.04415}
  {10.48550/arXiv.2312.04415} (\bibinfo {year} {2023}),\ \bibinfo {note}
  {arXiv:2312.04415 [cond-mat] type: article}\BibitemShut {NoStop}%
\bibitem [{\citenamefont {Gupta}\ \emph {et~al.}(2023)\citenamefont {Gupta},
  \citenamefont {Graziano}, \citenamefont {Pendharkar}, \citenamefont {Dong},
  \citenamefont {Dempsey}, \citenamefont {Palmstrøm},\ and\ \citenamefont
  {Pribiag}}]{gupta2023_diode}%
  \BibitemOpen
  \bibfield  {author} {\bibinfo {author} {\bibfnamefont {M.}~\bibnamefont
  {Gupta}}, \bibinfo {author} {\bibfnamefont {G.~V.}\ \bibnamefont {Graziano}},
  \bibinfo {author} {\bibfnamefont {M.}~\bibnamefont {Pendharkar}}, \bibinfo
  {author} {\bibfnamefont {J.~T.}\ \bibnamefont {Dong}}, \bibinfo {author}
  {\bibfnamefont {C.~P.}\ \bibnamefont {Dempsey}}, \bibinfo {author}
  {\bibfnamefont {C.}~\bibnamefont {Palmstrøm}},\ and\ \bibinfo {author}
  {\bibfnamefont {V.~S.}\ \bibnamefont {Pribiag}},\ }\href
  {https://doi.org/10.1038/s41467-023-38856-0} {\bibfield  {journal} {\bibinfo
  {journal} {Nature Communications}\ }\textbf {\bibinfo {volume} {14}},\
  \bibinfo {pages} {3078} (\bibinfo {year} {2023})},\ \bibinfo {note}
  {arXiv:2206.08471 [cond-mat] type: article}\BibitemShut {NoStop}%
\bibitem [{\citenamefont {Banerjee}\ \emph {et~al.}(2023)\citenamefont
  {Banerjee}, \citenamefont {Geier}, \citenamefont {Rahman}, \citenamefont
  {Thomas}, \citenamefont {Wang}, \citenamefont {Manfra}, \citenamefont
  {Flensberg},\ and\ \citenamefont {Marcus}}]{banerjee2023_diode}%
  \BibitemOpen
  \bibfield  {author} {\bibinfo {author} {\bibfnamefont {A.}~\bibnamefont
  {Banerjee}}, \bibinfo {author} {\bibfnamefont {M.}~\bibnamefont {Geier}},
  \bibinfo {author} {\bibfnamefont {M.~A.}\ \bibnamefont {Rahman}}, \bibinfo
  {author} {\bibfnamefont {C.}~\bibnamefont {Thomas}}, \bibinfo {author}
  {\bibfnamefont {T.}~\bibnamefont {Wang}}, \bibinfo {author} {\bibfnamefont
  {M.~J.}\ \bibnamefont {Manfra}}, \bibinfo {author} {\bibfnamefont
  {K.}~\bibnamefont {Flensberg}},\ and\ \bibinfo {author} {\bibfnamefont
  {C.~M.}\ \bibnamefont {Marcus}},\ }\href
  {https://doi.org/10.1103/PhysRevLett.131.196301} {\bibfield  {journal}
  {\bibinfo  {journal} {Physical Review Letters}\ }\textbf {\bibinfo {volume}
  {131}},\ \bibinfo {pages} {196301} (\bibinfo {year} {2023})}\BibitemShut
  {NoStop}%
\bibitem [{\citenamefont {Reinhardt}\ \emph {et~al.}(2024)\citenamefont
  {Reinhardt}, \citenamefont {Ascherl}, \citenamefont {Costa}, \citenamefont
  {Berger}, \citenamefont {Gronin}, \citenamefont {Gardner}, \citenamefont
  {Lindemann}, \citenamefont {Manfra}, \citenamefont {Fabian}, \citenamefont
  {Kochan}, \citenamefont {Strunk},\ and\ \citenamefont
  {Paradiso}}]{reinhardt2024}%
  \BibitemOpen
  \bibfield  {author} {\bibinfo {author} {\bibfnamefont {S.}~\bibnamefont
  {Reinhardt}}, \bibinfo {author} {\bibfnamefont {T.}~\bibnamefont {Ascherl}},
  \bibinfo {author} {\bibfnamefont {A.}~\bibnamefont {Costa}}, \bibinfo
  {author} {\bibfnamefont {J.}~\bibnamefont {Berger}}, \bibinfo {author}
  {\bibfnamefont {S.}~\bibnamefont {Gronin}}, \bibinfo {author} {\bibfnamefont
  {G.~C.}\ \bibnamefont {Gardner}}, \bibinfo {author} {\bibfnamefont
  {T.}~\bibnamefont {Lindemann}}, \bibinfo {author} {\bibfnamefont {M.~J.}\
  \bibnamefont {Manfra}}, \bibinfo {author} {\bibfnamefont {J.}~\bibnamefont
  {Fabian}}, \bibinfo {author} {\bibfnamefont {D.}~\bibnamefont {Kochan}},
  \bibinfo {author} {\bibfnamefont {C.}~\bibnamefont {Strunk}},\ and\ \bibinfo
  {author} {\bibfnamefont {N.}~\bibnamefont {Paradiso}},\ }\href
  {https://doi.org/10.1038/s41467-024-48741-z} {\bibfield  {journal} {\bibinfo
  {journal} {Nature Communications}\ }\textbf {\bibinfo {volume} {15}},\
  \bibinfo {pages} {4413} (\bibinfo {year} {2024})},\ \bibinfo {note}
  {arXiv:2308.01061 [cond-mat]}\BibitemShut {NoStop}%
\bibitem [{\citenamefont {Yuan}\ and\ \citenamefont {Fu}(2022)}]{yuan&fu2022}%
  \BibitemOpen
  \bibfield  {author} {\bibinfo {author} {\bibfnamefont {N.~F.~Q.}\
  \bibnamefont {Yuan}}\ and\ \bibinfo {author} {\bibfnamefont {L.}~\bibnamefont
  {Fu}},\ }\href {https://doi.org/10.1073/pnas.2119548119} {\bibfield
  {journal} {\bibinfo  {journal} {Proceedings of the National Academy of
  Sciences}\ }\textbf {\bibinfo {volume} {119}},\ \bibinfo {pages}
  {e2119548119} (\bibinfo {year} {2022})}\BibitemShut {NoStop}%
\bibitem [{\citenamefont {Ili{\'{c}}}\ and\ \citenamefont
  {Bergeret}(2022)}]{ilic&bergeret2022}%
  \BibitemOpen
  \bibfield  {author} {\bibinfo {author} {\bibfnamefont {S.}~\bibnamefont
  {Ili{\'{c}}}}\ and\ \bibinfo {author} {\bibfnamefont {F.~S.}\ \bibnamefont
  {Bergeret}},\ }\href {https://doi.org/10.1103/PhysRevLett.128.177001}
  {\bibfield  {journal} {\bibinfo  {journal} {Physical Review Letters}\
  }\textbf {\bibinfo {volume} {128}},\ \bibinfo {pages} {177001} (\bibinfo
  {year} {2022})}\BibitemShut {NoStop}%
\bibitem [{\citenamefont {Costa}\ \emph
  {et~al.}(2023{\natexlab{b}})\citenamefont {Costa}, \citenamefont {Fabian},\
  and\ \citenamefont {Kochan}}]{costa2023_diodeMicroscopic}%
  \BibitemOpen
  \bibfield  {author} {\bibinfo {author} {\bibfnamefont {A.}~\bibnamefont
  {Costa}}, \bibinfo {author} {\bibfnamefont {J.}~\bibnamefont {Fabian}},\ and\
  \bibinfo {author} {\bibfnamefont {D.}~\bibnamefont {Kochan}},\ }\href
  {https://doi.org/10.1103/PhysRevB.108.054522} {\bibfield  {journal} {\bibinfo
   {journal} {Physical Review B}\ }\textbf {\bibinfo {volume} {108}},\ \bibinfo
  {pages} {054522} (\bibinfo {year} {2023}{\natexlab{b}})},\ \bibinfo {note}
  {arXiv:2303.14823 [cond-mat] type: article}\BibitemShut {NoStop}%
\bibitem [{\citenamefont {Davydova}\ \emph {et~al.}(2022)\citenamefont
  {Davydova}, \citenamefont {Prembabu},\ and\ \citenamefont
  {Fu}}]{davydova2022}%
  \BibitemOpen
  \bibfield  {author} {\bibinfo {author} {\bibfnamefont {M.}~\bibnamefont
  {Davydova}}, \bibinfo {author} {\bibfnamefont {S.}~\bibnamefont {Prembabu}},\
  and\ \bibinfo {author} {\bibfnamefont {L.}~\bibnamefont {Fu}},\ }\href
  {https://doi.org/10.1126/sciadv.abo0309} {\bibfield  {journal} {\bibinfo
  {journal} {Science Advances}\ }\textbf {\bibinfo {volume} {8}},\ \bibinfo
  {pages} {eabo0309} (\bibinfo {year} {2022})}\BibitemShut {NoStop}%
\bibitem [{\citenamefont {Nakamura}\ \emph {et~al.}(2024)\citenamefont
  {Nakamura}, \citenamefont {Daido},\ and\ \citenamefont
  {Yanase}}]{nakamura2024}%
  \BibitemOpen
  \bibfield  {author} {\bibinfo {author} {\bibfnamefont {K.}~\bibnamefont
  {Nakamura}}, \bibinfo {author} {\bibfnamefont {A.}~\bibnamefont {Daido}},\
  and\ \bibinfo {author} {\bibfnamefont {Y.}~\bibnamefont {Yanase}},\ }\href
  {https://doi.org/10.1103/PhysRevB.109.094501} {\bibfield  {journal} {\bibinfo
   {journal} {Physical Review B}\ }\textbf {\bibinfo {volume} {109}},\ \bibinfo
  {pages} {094501} (\bibinfo {year} {2024})}\BibitemShut {NoStop}%
\bibitem [{\citenamefont {Pal}\ \emph {et~al.}(2022)\citenamefont {Pal},
  \citenamefont {Chakraborty}, \citenamefont {Sivakumar}, \citenamefont
  {Davydova}, \citenamefont {Gopi}, \citenamefont {Pandeya}, \citenamefont
  {Krieger}, \citenamefont {Zhang}, \citenamefont {Date}, \citenamefont {Ju},
  \citenamefont {Yuan}, \citenamefont {Schröter}, \citenamefont {Fu},\ and\
  \citenamefont {Parkin}}]{pal2022_diode}%
  \BibitemOpen
  \bibfield  {author} {\bibinfo {author} {\bibfnamefont {B.}~\bibnamefont
  {Pal}}, \bibinfo {author} {\bibfnamefont {A.}~\bibnamefont {Chakraborty}},
  \bibinfo {author} {\bibfnamefont {P.~K.}\ \bibnamefont {Sivakumar}}, \bibinfo
  {author} {\bibfnamefont {M.}~\bibnamefont {Davydova}}, \bibinfo {author}
  {\bibfnamefont {A.~K.}\ \bibnamefont {Gopi}}, \bibinfo {author}
  {\bibfnamefont {A.~K.}\ \bibnamefont {Pandeya}}, \bibinfo {author}
  {\bibfnamefont {J.~A.}\ \bibnamefont {Krieger}}, \bibinfo {author}
  {\bibfnamefont {Y.}~\bibnamefont {Zhang}}, \bibinfo {author} {\bibfnamefont
  {M.}~\bibnamefont {Date}}, \bibinfo {author} {\bibfnamefont {S.}~\bibnamefont
  {Ju}}, \bibinfo {author} {\bibfnamefont {N.}~\bibnamefont {Yuan}}, \bibinfo
  {author} {\bibfnamefont {N.~B.~M.}\ \bibnamefont {Schröter}}, \bibinfo
  {author} {\bibfnamefont {L.}~\bibnamefont {Fu}},\ and\ \bibinfo {author}
  {\bibfnamefont {S.~S.~P.}\ \bibnamefont {Parkin}},\ }\href
  {https://doi.org/10.1038/s41567-022-01699-5} {\bibfield  {journal} {\bibinfo
  {journal} {Nature Physics}\ }\textbf {\bibinfo {volume} {18}},\ \bibinfo
  {pages} {1228} (\bibinfo {year} {2022})}\BibitemShut {NoStop}%
\bibitem [{\citenamefont {Souto}\ \emph {et~al.}(2022)\citenamefont {Souto},
  \citenamefont {Leijnse},\ and\ \citenamefont {Schrade}}]{souto2022}%
  \BibitemOpen
  \bibfield  {author} {\bibinfo {author} {\bibfnamefont {R.~S.}\ \bibnamefont
  {Souto}}, \bibinfo {author} {\bibfnamefont {M.}~\bibnamefont {Leijnse}},\
  and\ \bibinfo {author} {\bibfnamefont {C.}~\bibnamefont {Schrade}},\ }\href
  {https://doi.org/10.1103/PhysRevLett.129.267702} {\bibfield  {journal}
  {\bibinfo  {journal} {Physical Review Letters}\ }\textbf {\bibinfo {volume}
  {129}},\ \bibinfo {pages} {267702} (\bibinfo {year} {2022})}\BibitemShut
  {NoStop}%
\bibitem [{\citenamefont {{{\v{Z}}}utic}\ \emph {et~al.}(2004)\citenamefont
  {{{\v{Z}}}utic}, \citenamefont {Fabian},\ and\ \citenamefont
  {Das~Sarma}}]{zutic2004review}%
  \BibitemOpen
  \bibfield  {author} {\bibinfo {author} {\bibfnamefont {I.}~\bibnamefont
  {{{\v{Z}}}utic}}, \bibinfo {author} {\bibfnamefont {J.}~\bibnamefont
  {Fabian}},\ and\ \bibinfo {author} {\bibfnamefont {S.}~\bibnamefont
  {Das~Sarma}},\ }\href {https://doi.org/10.1103/RevModPhys.76.323} {\bibfield
  {journal} {\bibinfo  {journal} {Reviews of Modern Physics}\ }\textbf
  {\bibinfo {volume} {76}},\ \bibinfo {pages} {323} (\bibinfo {year}
  {2004})}\BibitemShut {NoStop}%
\bibitem [{\citenamefont {Fabian}\ \emph {et~al.}(2007)\citenamefont {Fabian},
  \citenamefont {Matos-Abiague}, \citenamefont {Ertler}, \citenamefont
  {Stano},\ and\ \citenamefont {Zutic}}]{fabian2007review}%
  \BibitemOpen
  \bibfield  {author} {\bibinfo {author} {\bibfnamefont {J.}~\bibnamefont
  {Fabian}}, \bibinfo {author} {\bibfnamefont {A.}~\bibnamefont
  {Matos-Abiague}}, \bibinfo {author} {\bibfnamefont {C.}~\bibnamefont
  {Ertler}}, \bibinfo {author} {\bibfnamefont {P.}~\bibnamefont {Stano}},\ and\
  \bibinfo {author} {\bibfnamefont {I.}~\bibnamefont {Zutic}},\ }\bibfield
  {journal} {\bibinfo  {journal} {Acta Physica Slovaca. Reviews and Tutorials}\
  }\textbf {\bibinfo {volume} {57}},\ \href
  {https://doi.org/10.2478/v10155-010-0086-8} {10.2478/v10155-010-0086-8}
  (\bibinfo {year} {2007})\BibitemShut {NoStop}%
\bibitem [{\citenamefont {Schiela}\ \emph {et~al.}(2024)\citenamefont
  {Schiela}, \citenamefont {Yu},\ and\ \citenamefont
  {Shabani}}]{schiela2024perspective}%
  \BibitemOpen
  \bibfield  {author} {\bibinfo {author} {\bibfnamefont {W.~F.}\ \bibnamefont
  {Schiela}}, \bibinfo {author} {\bibfnamefont {P.}~\bibnamefont {Yu}},\ and\
  \bibinfo {author} {\bibfnamefont {J.}~\bibnamefont {Shabani}},\ }\href
  {https://doi.org/10.1103/PRXQuantum.5.030102} {\bibfield  {journal} {\bibinfo
   {journal} {PRX Quantum}\ }\textbf {\bibinfo {volume} {5}},\ \bibinfo {pages}
  {030102} (\bibinfo {year} {2024})}\BibitemShut {NoStop}%
\bibitem [{\citenamefont {Flensberg}\ \emph {et~al.}(2021)\citenamefont
  {Flensberg}, \citenamefont {von Oppen},\ and\ \citenamefont
  {Stern}}]{flensberg2021review}%
  \BibitemOpen
  \bibfield  {author} {\bibinfo {author} {\bibfnamefont {K.}~\bibnamefont
  {Flensberg}}, \bibinfo {author} {\bibfnamefont {F.}~\bibnamefont {von
  Oppen}},\ and\ \bibinfo {author} {\bibfnamefont {A.}~\bibnamefont {Stern}},\
  }\href {https://doi.org/10.1038/s41578-021-00336-6} {\bibfield  {journal}
  {\bibinfo  {journal} {Nature Reviews Materials}\ }\textbf {\bibinfo {volume}
  {6}},\ \bibinfo {pages} {944} (\bibinfo {year} {2021})}\BibitemShut {NoStop}%
\bibitem [{\citenamefont {Lutchyn}\ \emph {et~al.}(2018)\citenamefont
  {Lutchyn}, \citenamefont {Bakkers}, \citenamefont {Kouwenhoven},
  \citenamefont {Krogstrup}, \citenamefont {Marcus},\ and\ \citenamefont
  {Oreg}}]{lutchyn2018review}%
  \BibitemOpen
  \bibfield  {author} {\bibinfo {author} {\bibfnamefont {R.~M.}\ \bibnamefont
  {Lutchyn}}, \bibinfo {author} {\bibfnamefont {E.~P. A.~M.}\ \bibnamefont
  {Bakkers}}, \bibinfo {author} {\bibfnamefont {L.~P.}\ \bibnamefont
  {Kouwenhoven}}, \bibinfo {author} {\bibfnamefont {P.}~\bibnamefont
  {Krogstrup}}, \bibinfo {author} {\bibfnamefont {C.~M.}\ \bibnamefont
  {Marcus}},\ and\ \bibinfo {author} {\bibfnamefont {Y.}~\bibnamefont {Oreg}},\
  }\href {https://doi.org/10.1038/s41578-018-0003-1} {\bibfield  {journal}
  {\bibinfo  {journal} {Nature Reviews Materials}\ }\textbf {\bibinfo {volume}
  {3}},\ \bibinfo {pages} {52} (\bibinfo {year} {2018})}\BibitemShut {NoStop}%
\bibitem [{\citenamefont {Hell}\ \emph {et~al.}(2017)\citenamefont {Hell},
  \citenamefont {Leijnse},\ and\ \citenamefont
  {Flensberg}}]{hell2017_planarJJ}%
  \BibitemOpen
  \bibfield  {author} {\bibinfo {author} {\bibfnamefont {M.}~\bibnamefont
  {Hell}}, \bibinfo {author} {\bibfnamefont {M.}~\bibnamefont {Leijnse}},\ and\
  \bibinfo {author} {\bibfnamefont {K.}~\bibnamefont {Flensberg}},\ }\href
  {https://doi.org/10.1103/PhysRevLett.118.107701} {\bibfield  {journal}
  {\bibinfo  {journal} {Phys. Rev. Lett.}\ }\textbf {\bibinfo {volume} {118}},\
  \bibinfo {pages} {107701} (\bibinfo {year} {2017})}\BibitemShut {NoStop}%
\bibitem [{\citenamefont {Pientka}\ \emph {et~al.}(2017)\citenamefont
  {Pientka}, \citenamefont {Keselman}, \citenamefont {Berg}, \citenamefont
  {Yacoby}, \citenamefont {Stern},\ and\ \citenamefont
  {Halperin}}]{pientka2017}%
  \BibitemOpen
  \bibfield  {author} {\bibinfo {author} {\bibfnamefont {F.}~\bibnamefont
  {Pientka}}, \bibinfo {author} {\bibfnamefont {A.}~\bibnamefont {Keselman}},
  \bibinfo {author} {\bibfnamefont {E.}~\bibnamefont {Berg}}, \bibinfo {author}
  {\bibfnamefont {A.}~\bibnamefont {Yacoby}}, \bibinfo {author} {\bibfnamefont
  {A.}~\bibnamefont {Stern}},\ and\ \bibinfo {author} {\bibfnamefont {B.~I.}\
  \bibnamefont {Halperin}},\ }\href {https://doi.org/10.1103/PhysRevX.7.021032}
  {\bibfield  {journal} {\bibinfo  {journal} {Phys. Rev. X}\ }\textbf {\bibinfo
  {volume} {7}},\ \bibinfo {pages} {021032} (\bibinfo {year}
  {2017})}\BibitemShut {NoStop}%
\bibitem [{\citenamefont {Setiawan}\ \emph {et~al.}(2019)\citenamefont
  {Setiawan}, \citenamefont {Stern},\ and\ \citenamefont
  {Berg}}]{setiawan2019_narrowing}%
  \BibitemOpen
  \bibfield  {author} {\bibinfo {author} {\bibfnamefont {F.}~\bibnamefont
  {Setiawan}}, \bibinfo {author} {\bibfnamefont {A.}~\bibnamefont {Stern}},\
  and\ \bibinfo {author} {\bibfnamefont {E.}~\bibnamefont {Berg}},\ }\href
  {https://doi.org/10.1103/PhysRevB.99.220506} {\bibfield  {journal} {\bibinfo
  {journal} {Physical Review B}\ }\textbf {\bibinfo {volume} {99}},\ \bibinfo
  {pages} {220506} (\bibinfo {year} {2019})}\BibitemShut {NoStop}%
\bibitem [{\citenamefont {Haxell}\ \emph
  {et~al.}(2023{\natexlab{a}})\citenamefont {Haxell}, \citenamefont {Coraiola},
  \citenamefont {Sabonis}, \citenamefont {Hinderling}, \citenamefont {ten
  Kate}, \citenamefont {Cheah}, \citenamefont {Krizek}, \citenamefont {Schott},
  \citenamefont {Wegscheider},\ and\ \citenamefont
  {Nichele}}]{haxell2023_orbital}%
  \BibitemOpen
  \bibfield  {author} {\bibinfo {author} {\bibfnamefont {D.~Z.}\ \bibnamefont
  {Haxell}}, \bibinfo {author} {\bibfnamefont {M.}~\bibnamefont {Coraiola}},
  \bibinfo {author} {\bibfnamefont {D.}~\bibnamefont {Sabonis}}, \bibinfo
  {author} {\bibfnamefont {M.}~\bibnamefont {Hinderling}}, \bibinfo {author}
  {\bibfnamefont {S.~C.}\ \bibnamefont {ten Kate}}, \bibinfo {author}
  {\bibfnamefont {E.}~\bibnamefont {Cheah}}, \bibinfo {author} {\bibfnamefont
  {F.}~\bibnamefont {Krizek}}, \bibinfo {author} {\bibfnamefont
  {R.}~\bibnamefont {Schott}}, \bibinfo {author} {\bibfnamefont
  {W.}~\bibnamefont {Wegscheider}},\ and\ \bibinfo {author} {\bibfnamefont
  {F.}~\bibnamefont {Nichele}},\ }\bibfield  {journal} {\bibinfo  {journal}
  {ACS Nano}\ }\href {https://doi.org/10.1021/acsnano.3c04957}
  {10.1021/acsnano.3c04957} (\bibinfo {year} {2023}{\natexlab{a}})\BibitemShut
  {NoStop}%
\bibitem [{\citenamefont {Pekerten}\ \emph {et~al.}(2024)\citenamefont
  {Pekerten}, \citenamefont {Brandão}, \citenamefont {Bussiere}, \citenamefont
  {Monroe}, \citenamefont {Zhou}, \citenamefont {Han}, \citenamefont {Shabani},
  \citenamefont {Matos-Abiague},\ and\ \citenamefont
  {Žutić}}]{pekerten2024_beyondTopoJJStandardModel}%
  \BibitemOpen
  \bibfield  {author} {\bibinfo {author} {\bibfnamefont {B.}~\bibnamefont
  {Pekerten}}, \bibinfo {author} {\bibfnamefont {D.}~\bibnamefont {Brandão}},
  \bibinfo {author} {\bibfnamefont {B.}~\bibnamefont {Bussiere}}, \bibinfo
  {author} {\bibfnamefont {D.}~\bibnamefont {Monroe}}, \bibinfo {author}
  {\bibfnamefont {T.}~\bibnamefont {Zhou}}, \bibinfo {author} {\bibfnamefont
  {J.~E.}\ \bibnamefont {Han}}, \bibinfo {author} {\bibfnamefont
  {J.}~\bibnamefont {Shabani}}, \bibinfo {author} {\bibfnamefont
  {A.}~\bibnamefont {Matos-Abiague}},\ and\ \bibinfo {author} {\bibfnamefont
  {I.}~\bibnamefont {Žutić}},\ }\href {https://doi.org/10.1063/5.0214920}
  {\bibfield  {journal} {\bibinfo  {journal} {Applied Physics Letters}\
  }\textbf {\bibinfo {volume} {124}},\ \bibinfo {pages} {252602} (\bibinfo
  {year} {2024})},\ \bibinfo {note} {arXiv:2406.05829 [cond-mat]}\BibitemShut
  {NoStop}%
\bibitem [{\citenamefont {Fornieri}\ \emph {et~al.}(2019)\citenamefont
  {Fornieri}, \citenamefont {Whiticar}, \citenamefont {Setiawan}, \citenamefont
  {Portol{\'{e}}s}, \citenamefont {Drachmann}, \citenamefont {Keselman},
  \citenamefont {Gronin}, \citenamefont {Thomas}, \citenamefont {Wang},
  \citenamefont {Kallaher}, \citenamefont {Gardner}, \citenamefont {Berg},
  \citenamefont {Manfra}, \citenamefont {Stern}, \citenamefont {Marcus},\ and\
  \citenamefont {Nichele}}]{fornieri2019}%
  \BibitemOpen
  \bibfield  {author} {\bibinfo {author} {\bibfnamefont {A.}~\bibnamefont
  {Fornieri}}, \bibinfo {author} {\bibfnamefont {A.~M.}\ \bibnamefont
  {Whiticar}}, \bibinfo {author} {\bibfnamefont {F.}~\bibnamefont {Setiawan}},
  \bibinfo {author} {\bibfnamefont {E.}~\bibnamefont {Portol{\'{e}}s}},
  \bibinfo {author} {\bibfnamefont {A.~C.~C.}\ \bibnamefont {Drachmann}},
  \bibinfo {author} {\bibfnamefont {A.}~\bibnamefont {Keselman}}, \bibinfo
  {author} {\bibfnamefont {S.}~\bibnamefont {Gronin}}, \bibinfo {author}
  {\bibfnamefont {C.}~\bibnamefont {Thomas}}, \bibinfo {author} {\bibfnamefont
  {T.}~\bibnamefont {Wang}}, \bibinfo {author} {\bibfnamefont {R.}~\bibnamefont
  {Kallaher}}, \bibinfo {author} {\bibfnamefont {G.~C.}\ \bibnamefont
  {Gardner}}, \bibinfo {author} {\bibfnamefont {E.}~\bibnamefont {Berg}},
  \bibinfo {author} {\bibfnamefont {M.~J.}\ \bibnamefont {Manfra}}, \bibinfo
  {author} {\bibfnamefont {A.}~\bibnamefont {Stern}}, \bibinfo {author}
  {\bibfnamefont {C.~M.}\ \bibnamefont {Marcus}},\ and\ \bibinfo {author}
  {\bibfnamefont {F.}~\bibnamefont {Nichele}},\ }\href
  {https://doi.org/10.1038/s41586-019-1068-8} {\bibfield  {journal} {\bibinfo
  {journal} {Nature}\ }\textbf {\bibinfo {volume} {569}},\ \bibinfo {pages}
  {89} (\bibinfo {year} {2019})}\BibitemShut {NoStop}%
\bibitem [{\citenamefont {Dartiailh}\ \emph {et~al.}(2021)\citenamefont
  {Dartiailh}, \citenamefont {Mayer}, \citenamefont {Yuan}, \citenamefont
  {Wickramasinghe}, \citenamefont {Matos-Abiague}, \citenamefont {\ifmmode
  \check{Z}\else \v{Z}\fi{}uti\ifmmode~\acute{c}\else \'{c}\fi{}},\ and\
  \citenamefont {Shabani}}]{dartiailh2021_piJump}%
  \BibitemOpen
  \bibfield  {author} {\bibinfo {author} {\bibfnamefont {M.~C.}\ \bibnamefont
  {Dartiailh}}, \bibinfo {author} {\bibfnamefont {W.}~\bibnamefont {Mayer}},
  \bibinfo {author} {\bibfnamefont {J.}~\bibnamefont {Yuan}}, \bibinfo {author}
  {\bibfnamefont {K.~S.}\ \bibnamefont {Wickramasinghe}}, \bibinfo {author}
  {\bibfnamefont {A.}~\bibnamefont {Matos-Abiague}}, \bibinfo {author}
  {\bibfnamefont {I.}~\bibnamefont {\ifmmode \check{Z}\else
  \v{Z}\fi{}uti\ifmmode~\acute{c}\else \'{c}\fi{}}},\ and\ \bibinfo {author}
  {\bibfnamefont {J.}~\bibnamefont {Shabani}},\ }\href
  {https://doi.org/10.1103/PhysRevLett.126.036802} {\bibfield  {journal}
  {\bibinfo  {journal} {Phys. Rev. Lett.}\ }\textbf {\bibinfo {volume} {126}},\
  \bibinfo {pages} {036802} (\bibinfo {year} {2021})}\BibitemShut {NoStop}%
\bibitem [{\citenamefont {Ren}\ \emph {et~al.}(2019)\citenamefont {Ren},
  \citenamefont {Pientka}, \citenamefont {Hart}, \citenamefont {Pierce},
  \citenamefont {Kosowsky}, \citenamefont {Lunczer}, \citenamefont {Schlereth},
  \citenamefont {Scharf}, \citenamefont {Hankiewicz}, \citenamefont
  {Molenkamp}, \citenamefont {Halperin},\ and\ \citenamefont
  {Yacoby}}]{ren2019}%
  \BibitemOpen
  \bibfield  {author} {\bibinfo {author} {\bibfnamefont {H.}~\bibnamefont
  {Ren}}, \bibinfo {author} {\bibfnamefont {F.}~\bibnamefont {Pientka}},
  \bibinfo {author} {\bibfnamefont {S.}~\bibnamefont {Hart}}, \bibinfo {author}
  {\bibfnamefont {A.~T.}\ \bibnamefont {Pierce}}, \bibinfo {author}
  {\bibfnamefont {M.}~\bibnamefont {Kosowsky}}, \bibinfo {author}
  {\bibfnamefont {L.}~\bibnamefont {Lunczer}}, \bibinfo {author} {\bibfnamefont
  {R.}~\bibnamefont {Schlereth}}, \bibinfo {author} {\bibfnamefont
  {B.}~\bibnamefont {Scharf}}, \bibinfo {author} {\bibfnamefont {E.~M.}\
  \bibnamefont {Hankiewicz}}, \bibinfo {author} {\bibfnamefont {L.~W.}\
  \bibnamefont {Molenkamp}}, \bibinfo {author} {\bibfnamefont {B.~I.}\
  \bibnamefont {Halperin}},\ and\ \bibinfo {author} {\bibfnamefont
  {A.}~\bibnamefont {Yacoby}},\ }\href
  {https://doi.org/10.1038/s41586-019-1148-9} {\bibfield  {journal} {\bibinfo
  {journal} {Nature}\ }\textbf {\bibinfo {volume} {569}},\ \bibinfo {pages}
  {93} (\bibinfo {year} {2019})}\BibitemShut {NoStop}%
\bibitem [{\citenamefont {Meservey}\ and\ \citenamefont
  {Tedrow}(1971)}]{meservey&tedrow1971}%
  \BibitemOpen
  \bibfield  {author} {\bibinfo {author} {\bibfnamefont {R.}~\bibnamefont
  {Meservey}}\ and\ \bibinfo {author} {\bibfnamefont {P.~M.}\ \bibnamefont
  {Tedrow}},\ }\href {https://doi.org/10.1063/1.1659648} {\bibfield  {journal}
  {\bibinfo  {journal} {Journal of Applied Physics}\ }\textbf {\bibinfo
  {volume} {42}},\ \bibinfo {pages} {51} (\bibinfo {year} {1971})}\BibitemShut
  {NoStop}%
\bibitem [{\citenamefont {Nazarov}\ and\ \citenamefont
  {Blanter}(2009)}]{nazarov}%
  \BibitemOpen
  \bibfield  {author} {\bibinfo {author} {\bibfnamefont {Y.~V.}\ \bibnamefont
  {Nazarov}}\ and\ \bibinfo {author} {\bibfnamefont {Y.~M.}\ \bibnamefont
  {Blanter}},\ }\href {https://doi.org/10.1017/CBO9780511626906} {\emph
  {\bibinfo {title} {Quantum Transport: Introduction to Nanoscience}}}\
  (\bibinfo  {publisher} {Cambridge University Press},\ \bibinfo {year}
  {2009})\BibitemShut {NoStop}%
\bibitem [{\citenamefont {Prada}\ \emph {et~al.}(2020)\citenamefont {Prada},
  \citenamefont {San-Jose}, \citenamefont {de~Moor}, \citenamefont {Geresdi},
  \citenamefont {Lee}, \citenamefont {Klinovaja}, \citenamefont {Loss},
  \citenamefont {Nyg{{\aa}}rd}, \citenamefont {Aguado},\ and\ \citenamefont
  {Kouwenhoven}}]{prada2020review}%
  \BibitemOpen
  \bibfield  {author} {\bibinfo {author} {\bibfnamefont {E.}~\bibnamefont
  {Prada}}, \bibinfo {author} {\bibfnamefont {P.}~\bibnamefont {San-Jose}},
  \bibinfo {author} {\bibfnamefont {M.~W.~A.}\ \bibnamefont {de~Moor}},
  \bibinfo {author} {\bibfnamefont {A.}~\bibnamefont {Geresdi}}, \bibinfo
  {author} {\bibfnamefont {E.~J.~H.}\ \bibnamefont {Lee}}, \bibinfo {author}
  {\bibfnamefont {J.}~\bibnamefont {Klinovaja}}, \bibinfo {author}
  {\bibfnamefont {D.}~\bibnamefont {Loss}}, \bibinfo {author} {\bibfnamefont
  {J.}~\bibnamefont {Nyg{{\aa}}rd}}, \bibinfo {author} {\bibfnamefont
  {R.}~\bibnamefont {Aguado}},\ and\ \bibinfo {author} {\bibfnamefont {L.~P.}\
  \bibnamefont {Kouwenhoven}},\ }\href
  {https://doi.org/10.1038/s42254-020-0228-y} {\bibfield  {journal} {\bibinfo
  {journal} {Nature Reviews Physics}\ }\textbf {\bibinfo {volume} {2}},\
  \bibinfo {pages} {575} (\bibinfo {year} {2020})}\BibitemShut {NoStop}%
\bibitem [{\citenamefont {Kulik}\ and\ \citenamefont
  {Omel'yanchuk}(1975)}]{kulik&omelyanchuk1975}%
  \BibitemOpen
  \bibfield  {author} {\bibinfo {author} {\bibfnamefont {I.~O.}\ \bibnamefont
  {Kulik}}\ and\ \bibinfo {author} {\bibfnamefont {A.~N.}\ \bibnamefont
  {Omel'yanchuk}},\ }\href {https://www.osti.gov/biblio/4209268} {\bibfield
  {journal} {\bibinfo  {journal} {JETP Lett.}\ }\textbf {\bibinfo {volume}
  {21}},\ \bibinfo {pages} {96} (\bibinfo {year} {1975})},\ \bibinfo {note}
  {zhETF Pis. Red. 21(4), 216--219 (1975)}\BibitemShut {NoStop}%
\bibitem [{\citenamefont {Mayer}\ \emph {et~al.}(2019)\citenamefont {Mayer},
  \citenamefont {Yuan}, \citenamefont {Wickramasinghe}, \citenamefont {Nguyen},
  \citenamefont {Dartiailh},\ and\ \citenamefont
  {Shabani}}]{mayer2019_proximitizedInAs}%
  \BibitemOpen
  \bibfield  {author} {\bibinfo {author} {\bibfnamefont {W.}~\bibnamefont
  {Mayer}}, \bibinfo {author} {\bibfnamefont {J.}~\bibnamefont {Yuan}},
  \bibinfo {author} {\bibfnamefont {K.~S.}\ \bibnamefont {Wickramasinghe}},
  \bibinfo {author} {\bibfnamefont {T.}~\bibnamefont {Nguyen}}, \bibinfo
  {author} {\bibfnamefont {M.~C.}\ \bibnamefont {Dartiailh}},\ and\ \bibinfo
  {author} {\bibfnamefont {J.}~\bibnamefont {Shabani}},\ }\href
  {https://doi.org/10.1063/1.5067363} {\bibfield  {journal} {\bibinfo
  {journal} {Applied Physics Letters}\ }\textbf {\bibinfo {volume} {114}},\
  \bibinfo {pages} {103104} (\bibinfo {year} {2019})}\BibitemShut {NoStop}%
\bibitem [{\citenamefont {Adachi}(2005)}]{adachi_semiconductors}%
  \BibitemOpen
  \bibfield  {author} {\bibinfo {author} {\bibfnamefont {S.}~\bibnamefont
  {Adachi}},\ }\href@noop {} {\emph {\bibinfo {title} {Properties of
  group-{IV}, {III-V}, and {II-VI} semiconductors}}},\ Wiley Series in
  Materials for Electronic \& Optoelectronic Applications\ (\bibinfo
  {publisher} {Wiley},\ \bibinfo {year} {2005})\BibitemShut {NoStop}%
\bibitem [{\citenamefont {Pöschl}\ \emph {et~al.}(2022)\citenamefont
  {Pöschl}, \citenamefont {Danilenko}, \citenamefont {Sabonis}, \citenamefont
  {Kristjuhan}, \citenamefont {Lindemann}, \citenamefont {Thomas},
  \citenamefont {Manfra},\ and\ \citenamefont {Marcus}}]{poschl2022_1}%
  \BibitemOpen
  \bibfield  {author} {\bibinfo {author} {\bibfnamefont {A.}~\bibnamefont
  {Pöschl}}, \bibinfo {author} {\bibfnamefont {A.}~\bibnamefont {Danilenko}},
  \bibinfo {author} {\bibfnamefont {D.}~\bibnamefont {Sabonis}}, \bibinfo
  {author} {\bibfnamefont {K.}~\bibnamefont {Kristjuhan}}, \bibinfo {author}
  {\bibfnamefont {T.}~\bibnamefont {Lindemann}}, \bibinfo {author}
  {\bibfnamefont {C.}~\bibnamefont {Thomas}}, \bibinfo {author} {\bibfnamefont
  {M.~J.}\ \bibnamefont {Manfra}},\ and\ \bibinfo {author} {\bibfnamefont
  {C.~M.}\ \bibnamefont {Marcus}},\ }\href
  {https://doi.org/10.1103/PhysRevB.106.L161301} {\bibfield  {journal}
  {\bibinfo  {journal} {Physical Review B}\ }\textbf {\bibinfo {volume}
  {106}},\ \bibinfo {pages} {L161301} (\bibinfo {year} {2022})}\BibitemShut
  {NoStop}%
\bibitem [{\citenamefont {Haxell}\ \emph
  {et~al.}(2023{\natexlab{b}})\citenamefont {Haxell}, \citenamefont {Coraiola},
  \citenamefont {Hinderling}, \citenamefont {ten Kate}, \citenamefont
  {Sabonis}, \citenamefont {Svetogorov}, \citenamefont {Belzig}, \citenamefont
  {Cheah}, \citenamefont {Krizek}, \citenamefont {Schott}, \citenamefont
  {Wegscheider},\ and\ \citenamefont {Nichele}}]{haxell2023_andreevMolecule}%
  \BibitemOpen
  \bibfield  {author} {\bibinfo {author} {\bibfnamefont {D.~Z.}\ \bibnamefont
  {Haxell}}, \bibinfo {author} {\bibfnamefont {M.}~\bibnamefont {Coraiola}},
  \bibinfo {author} {\bibfnamefont {M.}~\bibnamefont {Hinderling}}, \bibinfo
  {author} {\bibfnamefont {S.~C.}\ \bibnamefont {ten Kate}}, \bibinfo {author}
  {\bibfnamefont {D.}~\bibnamefont {Sabonis}}, \bibinfo {author} {\bibfnamefont
  {A.~E.}\ \bibnamefont {Svetogorov}}, \bibinfo {author} {\bibfnamefont
  {W.}~\bibnamefont {Belzig}}, \bibinfo {author} {\bibfnamefont
  {E.}~\bibnamefont {Cheah}}, \bibinfo {author} {\bibfnamefont
  {F.}~\bibnamefont {Krizek}}, \bibinfo {author} {\bibfnamefont
  {R.}~\bibnamefont {Schott}}, \bibinfo {author} {\bibfnamefont
  {W.}~\bibnamefont {Wegscheider}},\ and\ \bibinfo {author} {\bibfnamefont
  {F.}~\bibnamefont {Nichele}},\ }\href
  {https://doi.org/10.1021/acs.nanolett.3c02066} {\bibfield  {journal}
  {\bibinfo  {journal} {Nano Letters}\ }\textbf {\bibinfo {volume} {23}},\
  \bibinfo {pages} {7532} (\bibinfo {year} {2023}{\natexlab{b}})}\BibitemShut
  {NoStop}%
\bibitem [{\citenamefont {Farzaneh}\ \emph {et~al.}(2024)\citenamefont
  {Farzaneh}, \citenamefont {Hatefipour}, \citenamefont {Schiela},
  \citenamefont {Lotfizadeh}, \citenamefont {Yu}, \citenamefont {Elfeky},
  \citenamefont {Strickland}, \citenamefont {Matos-Abiague},\ and\
  \citenamefont {Shabani}}]{farzaneh2024}%
  \BibitemOpen
  \bibfield  {author} {\bibinfo {author} {\bibfnamefont {S.~M.}\ \bibnamefont
  {Farzaneh}}, \bibinfo {author} {\bibfnamefont {M.}~\bibnamefont
  {Hatefipour}}, \bibinfo {author} {\bibfnamefont {W.~F.}\ \bibnamefont
  {Schiela}}, \bibinfo {author} {\bibfnamefont {N.}~\bibnamefont {Lotfizadeh}},
  \bibinfo {author} {\bibfnamefont {P.}~\bibnamefont {Yu}}, \bibinfo {author}
  {\bibfnamefont {B.~H.}\ \bibnamefont {Elfeky}}, \bibinfo {author}
  {\bibfnamefont {W.~M.}\ \bibnamefont {Strickland}}, \bibinfo {author}
  {\bibfnamefont {A.}~\bibnamefont {Matos-Abiague}},\ and\ \bibinfo {author}
  {\bibfnamefont {J.}~\bibnamefont {Shabani}},\ }\href
  {https://doi.org/10.1103/PhysRevResearch.6.013039} {\bibfield  {journal}
  {\bibinfo  {journal} {Physical Review Research}\ }\textbf {\bibinfo {volume}
  {6}},\ \bibinfo {pages} {013039} (\bibinfo {year} {2024})}\BibitemShut
  {NoStop}%
\bibitem [{\citenamefont {Hart}\ \emph {et~al.}(2017)\citenamefont {Hart},
  \citenamefont {Ren}, \citenamefont {Kosowsky}, \citenamefont {Ben-Shach},
  \citenamefont {Leubner}, \citenamefont {Br{\"{u}}ne}, \citenamefont
  {Buhmann}, \citenamefont {Molenkamp}, \citenamefont {Halperin},\ and\
  \citenamefont {Yacoby}}]{hart2017}%
  \BibitemOpen
  \bibfield  {author} {\bibinfo {author} {\bibfnamefont {S.}~\bibnamefont
  {Hart}}, \bibinfo {author} {\bibfnamefont {H.}~\bibnamefont {Ren}}, \bibinfo
  {author} {\bibfnamefont {M.}~\bibnamefont {Kosowsky}}, \bibinfo {author}
  {\bibfnamefont {G.}~\bibnamefont {Ben-Shach}}, \bibinfo {author}
  {\bibfnamefont {P.}~\bibnamefont {Leubner}}, \bibinfo {author} {\bibfnamefont
  {C.}~\bibnamefont {Br{\"{u}}ne}}, \bibinfo {author} {\bibfnamefont
  {H.}~\bibnamefont {Buhmann}}, \bibinfo {author} {\bibfnamefont {L.~W.}\
  \bibnamefont {Molenkamp}}, \bibinfo {author} {\bibfnamefont {B.~I.}\
  \bibnamefont {Halperin}},\ and\ \bibinfo {author} {\bibfnamefont
  {A.}~\bibnamefont {Yacoby}},\ }\href {https://doi.org/10.1038/nphys3877}
  {\bibfield  {journal} {\bibinfo  {journal} {Nature Physics}\ }\textbf
  {\bibinfo {volume} {13}},\ \bibinfo {pages} {87} (\bibinfo {year}
  {2017})}\BibitemShut {NoStop}%
\bibitem [{\citenamefont {Scharf}\ \emph {et~al.}(2024)\citenamefont {Scharf},
  \citenamefont {Kochan},\ and\ \citenamefont {Matos-Abiague}}]{scharf2024}%
  \BibitemOpen
  \bibfield  {author} {\bibinfo {author} {\bibfnamefont {B.}~\bibnamefont
  {Scharf}}, \bibinfo {author} {\bibfnamefont {D.}~\bibnamefont {Kochan}},\
  and\ \bibinfo {author} {\bibfnamefont {A.}~\bibnamefont {Matos-Abiague}},\
  }\href {https://doi.org/10.1103/PhysRevB.110.134511} {\bibfield  {journal}
  {\bibinfo  {journal} {Physical Review B}\ }\textbf {\bibinfo {volume}
  {110}},\ \bibinfo {pages} {134511} (\bibinfo {year} {2024})}\BibitemShut
  {NoStop}%
\end{thebibliography}%
\end{document}